\documentclass{aa}  
\usepackage{graphicx}
\usepackage{txfonts}
\usepackage{natbib}
\usepackage{threeparttable}
\usepackage{caption}
\usepackage{subcaption}
\usepackage{color}
\usepackage[colorlinks=true,linkcolor=blue]{hyperref}

\usepackage{url}
\usepackage{tabularx}

\usepackage[normalem]{ulem}

\begin{document} 
\defcitealias{NK13}{N13}

%https://www.overleaf.com/project/5f48bd1a796c4d0001612eb3
%\title{The compactness of Ultra faint dwarf galaxies : a  new challenge for numerical simulations ?}

\title{The compactness of ultra-faint dwarf galaxies: A  new challenge? }

%\title{On the compactness of Ultra faint dwarf galaxies.}
   %\title{The metallicity plateau of ultra faint dwarf galaxies : the smoking gun of Pop III PISNe stars ?}
   %\title{The necessity of Pop III stars to fit the metallicity relation in the faintest galaxies}
   %\title{An indirect poof of the existence of Pop III stars in the faintest galaxies}

   \author{Yves Revaz\inst{\ref{epfl}}}

   \institute{Institute of Physics, Laboratory of Astrophysics, \'{E}cole Polytechnique F\'{e}d\'{e}rale de Lausanne (EPFL), 1290 Sauverny, Switzerland \label{epfl}}

   %\author{Yves Revaz\inst{\ref{epfl}}, 
   %        Mahsa Sanati\inst{\ref{epfl}},
   %        Nicolas Longeard\inst{\ref{epfl}},
   %        Pascale Jablonka\inst{\ref{epfl}, \ref{Observatoire de Paris}}
   %       }

   %\institute{Institute of Physics, Laboratory of Astrophysics, \'{E}cole Polytechnique F\'{e}d\'{e}rale de %Lausanne (EPFL), 1290 Sauverny, Switzerland \label{epfl}\\
   %     \email{mahsa.sanati@epfl.ch}
   % \and GEPI, CNRS UMR 8111, Observatoire de Paris, PSL University, 92125 Meudon, Cedex, France %\label{Observatoire de Paris}
   %          }

   \date{Received: XXXX; accepted: YYYY}
% \abstract{}{}{}{}{} 
% 5 {} token are mandatory
 
  \abstract
    { 
    %\textbf{Aim} 
    So far, numerical simulations of ultra-faint dwarf galaxies (UFDs) have failed to 
    properly reproduce the observed size--luminosity relation. 
    In particular, no hydrodynamical simulation run has managed to form UFDs with a half-light 
    radius as small as $30\,\rm{pc}$, as seen in observations of several UFD candidates.
    We tackle this problem by developing a simple but numerically clean and powerful 
    method in which predictions 
    of the stellar content of UFDs from $\Lambda$CDM cosmological 
    hydrodynamical simulations are combined with very high-resolution dark-matter-only runs.
    This method allows us to trace the buildup history of UFDs and to determine 
    the impact of the merger of building-block objects on their final size.    
    We find that, while no UFDs more compact than $20\,\rm{pc}$ can be formed,
    slightly larger systems are only reproduced if all member stars originate 
    from the same initial mini-halo. However, this imposes that 
    (i) the total virial mass is smaller than $3\cdot 10^8\,\rm{M_\odot}$,
    and (ii) the stellar content prior to the end of the reionisation epoch is very 
    compact ($<15\,\rm{pc}$) and strongly gravitationally bound, which is  
    a challenge for current hydrodynamical numerical simulations.
    If initial stellar building blocks are larger than $35\,\rm{pc,}$
    the size of the UFD will extend to $80\,\rm{pc}$.
    Finally, our study shows that UFDs keep strong imprints of their buildup history in the form of elongated or extended stellar halos. Those features can erroneously be interpreted as tidal signatures.

    %This study will help to desing 

    %While the difficulty in reproducing the compactness of the faintest UFDs 
    %could reveal some limitation of the cutting-edge numerical models, it could
    %also be a natural prediction of the hierarchical $\Lambda$CDM model.

    %This method allows us to study to which extend the complex build-up history of 
    %UFDs, as predicted by the $\Lambda$CDM paradigm, is responsible of the formation 
    %of over-sized UFDs in cosmological simulations.

%Using a simple but powerful and numerically clean method that combines predictions 
%of the stellar content of UFDs from hydro-simulations with DMO ones, we followed the 
%build-up history of UFDs 
    
   }


   \titlerunning{On the compactness of Ultra faint dwarf galaxies}   
   \authorrunning{Y. Revaz et al}

   \maketitle

%
%%%%%%%%%%%%%%%%%%%%%%%%%%%%%%%%%%%%%%%%%%%%%%%%%%%%%%%%%%%%%%%%

\section{Introduction}\label{sec:intro}

%%%%%%%%%%%%%%%%%%%%%%%%%%%%%%%%%%%%%%%%%%%%%%%%%%%%%%%%%%%%%%%%

%Il faut lire:
%https://ui.adsabs.harvard.edu/abs/2023arXiv230212818O/abstract
%en fait tout les papier edge avec Agertz

Ultra-faint dwarf galaxies (UFDs) are the faintest galaxies known in the Universe, with, by definition, a V-band luminosity of fainter than $L_\mathrm{V}$=$10^5\,L_{\odot}$, $M_{\rm{V}} < -7.7$ \citep[see][]{Simon_2019}.
Among them, some can be as faint as only a few hundred times solar, like Triangulum II
\citep{laevens2015,kirby2015},
Segue I
\citep{belokurov2007},
or Segue II
\citep{belokurov2009}.
% Old stellar populations
UFDs are dominated by very old, that is, $>10\,\rm{Gyr}$ \citep{okamoto2012,brown2012}, and metal-poor ($\rm{[Fe/H]}\cong -2\,\rm{to}\,-3$) \citep{munoz2006,martin2007,simon2007,kirby2008,fu2023} stellar populations.
Their structural properties and morphology have now been debated for over 15 years.
It has been quickly shown that UFDs are rather elongated systems
\citep{martin2008} with a mean ellipticity ($\epsilon$) of about $0.4$.
Such a relatively elongated shape has been postulated to result from tidal 
interaction with the Milky Way \citep{martin2008,sand2009,longeard2022}.
%
% Hercules
Among the most extreme systems, Hercules ($\epsilon=0.67$) 
seems to be extended by a stream of stars \citep{sand2009,roderick2015}, 
a sign that it is a strongly tidally disrupted system.
This conclusion is supported by tentative evidence of a 
velocity gradient \citep{aden2009}.
However, relying on three new star members, \citet{longeard2023}
were not able to confirm this gradient and concluded that Hercules 
does not show evidence of ongoing tidal disruption.
% Bootes I
Bo\"otes I ($\epsilon=0.68$) is another example of an elongated UFD. 
In addition to a velocity gradient, \citet{longeard2022} reveal the presence
of a metallicity gradient, both suggesting that Bo\"otes I may have also been affected 
by tides.
On the other hand, \citet{frebel2016} reported that Bo\"otes I 
also displays two stellar populations,  indicating that it could originate
from different building-block objects.
% Tucana II 
Another interesting morphological feature is
the extended stellar halo of the Tucana II galaxy \citep{chiti2021}
revealing either a strong bursty feedback process
or an early galactic merger.

% DM
With an average size of less than $500\,\rm{pc}$, UFDs are also the
smallest galaxies known. However, compared to globular clusters, they exhibit 
a larger velocity dispersion of between $2$ and nearly $10\,\rm{km/s}$
\citep{kleyna2005,munoz2006,martin2007,simon2007,martin2016,Simon_2019,longeard2023}
and a broader metallicity distribution.
Combining their size and kinematics, and assuming they are self-gravitating systems 
at equilibrium, this makes UFDs the most dark-matter-dominated galactic systems in our 
Universe. As such, they constitute fundamental probes of the cosmological model \citep{Bullock2017}, with the ability to constrain the total matter power spectrum 
down to the smallest scales \citep{sanati2020}.

% Reproduction of UFDs from numerical simulations
Several groups have studied their formation and evolution in a $\Lambda$CDM cosmological
context using numerical schemes that include a complex treatment of the baryonic physics  \citep{jeon2017,Wheeler_2019,agertz2020,applebaum2020,jeon2021a,jeon2021b,orkney2021,prgomet2022,gandhi2022,sanati2023}.
In these simulations, the rapid star formation quenching of UFDs is interpreted as a direct consequence 
of their very low potential well. Ionising photons emitted by young and
massive stars ---and responsible for the re-ionisation of the Universe during
the first billions years--- heat the cold and star forming gas expelled from the galaxy, causing the star formation to cease abruptly. 
UFDs are thus compatible with the first galaxies formed
in mini-halos before the epoch of reionisation (EoR)
\citep{Ricotti_2005,Wyithe_2006,salvadori2009,2009MNRAS.395L...6S,bovill2009,Wheeler_2019,Rodriguez2019,sanati2023}.

While their luminosity, velocity dispersion, and
star formation histories are relatively well reproduced by the aforementioned numerical simulations, some of their properties remain challenging
to understand.
One in particular is the metallicity distribution function and how it correlates with the luminosity through the metallicity--luminosity relation \citep{sanati2023,fu2023}.
Numerical models predict that UFDs will contain much less metal than is suggested by 
observations at a given luminosity and fail to reproduce the presence of
{relatively metal-rich} stars with $\rm{[Fe/H]} \cong -1$.
Some attempts to solve this problem have been
made by either assuming
a metallicity-dependent IMF \citep{prgomet2022},
a metallicity-dependent SNIa-rate \citep{gandhi2022},
or an enrichment by pair-instability supernovae \citep{sanati2023}.
However, none of these solutions can fully solve the problem.

In addition to the metallicity--luminosity relation,
the size--luminosity relation is also challenging.
As described in detail in Sect.~\ref{sec:challenge}, most 
hydrodynamical models over-predict the size of UFDs 
and fail to reproduce the most compact observed ones.

In this paper we address this latter problem by tracing 
the size evolution of collisionless initially compact UFDs formed at the end 
of the EoR. We study their evolution in the framework of the $\Lambda$CDM paradigm
where UFD buildup is predicted to be quite complex and will naturally affect their size.

The paper is organised as follows:
In Sect.~\ref{sec:challenge}, we provide a detailed review of the discrepancy between the observed and simulated luminosity--size relation. 
In Sect.~\ref{sec:simulations}, we describe the numerical methods
developed to study the size evolution of UFDs in a cosmological context.
Our results are presented in Sect.~\ref{sec:results}.
A discussion and conclusions are given 
in Sect.~\ref{sec:conclusions}.

%%%%%%%%%%%%%%%%%%%%%%%%%%%%%%%%%%%%%%%%%%%%%%%%%%%%%%%%%%%%%%%%

\section{The challenge of the luminosity--size relation}\label{sec:challenge}

%%%%%%%%%%%%%%%%%%%%%%%%%%%%%%%%%%%%%%%%%%%%%%%%%%%%%%%%%%%%%%%%

\begin{figure}[h]
  \centering
  \includegraphics[width=0.48\textwidth]{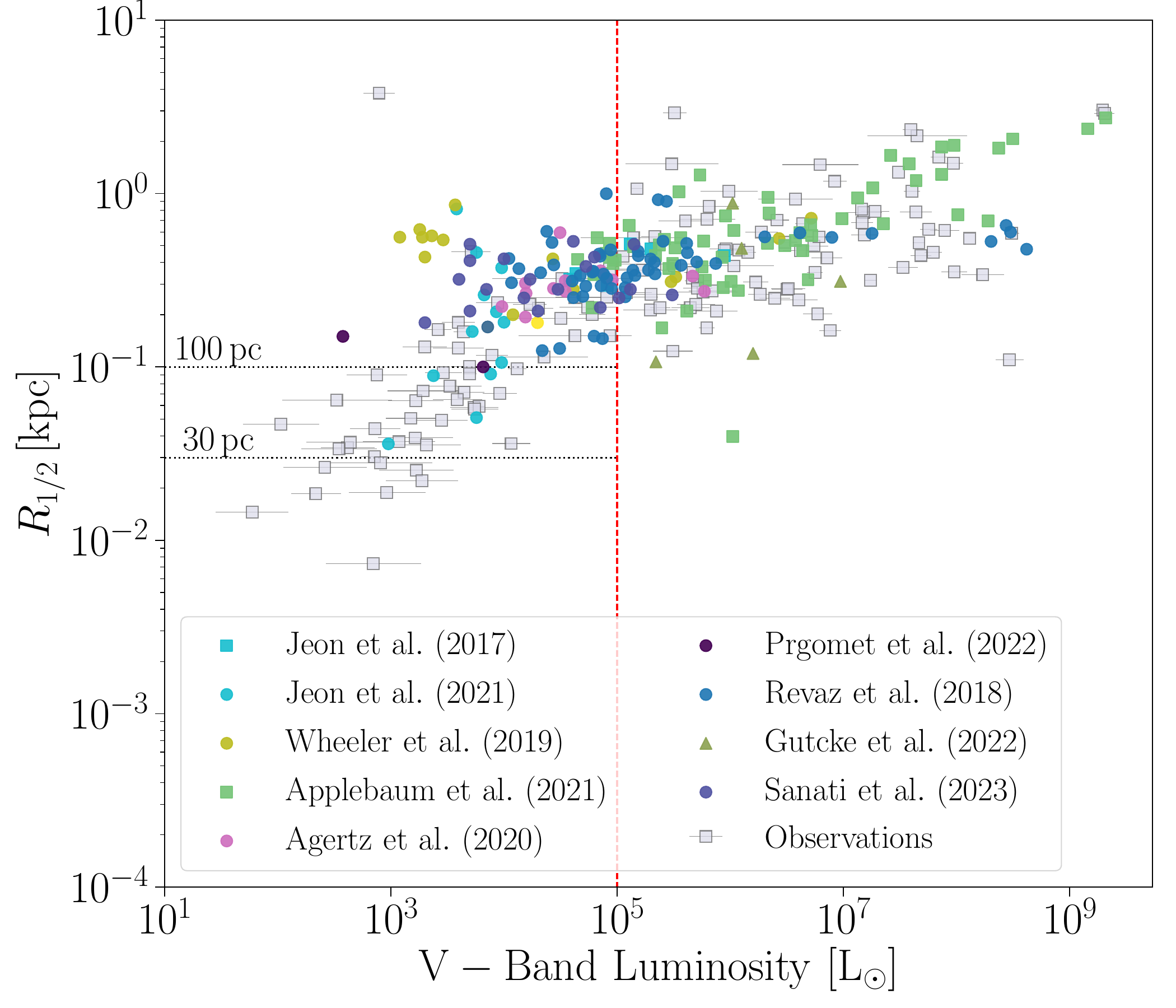}
  \caption{\small Comparison of the luminosity--size relation for dwarfs and UFDs as derived from both observations and simulations. The grey squares represent our UFD sample (see text for details). 
  Coloured points stand for different simulations published with the references given at the bottom of the figure. 
  The vertical red dotted line separates UFDs from brighter dwarfs.
  %We extended our own work \citep{Revaz2018} with the results of unpublished galaxies in the UFDs regime.
  \label{fig:r12L}
  }
  
\end{figure}

Similarly to more massive galaxies, dwarf galaxies follow different scaling relations. 
The size--luminosity relation shows that brighter dwarfs are more 
extended than fainter ones. 
While this correlation has been well known for dwarfs for many years now \citep{kormendy1985,McConnachie2012}, it
is only recently that this relation has been extended to UFD galaxies \citep{Simon_2019}.
In this relation, the size is conventionally represented by the 2D half-light radius $R_{1/2}$ (also
known as the effective radius), defined as the projected radius that contains half of the total luminosity
of the galaxy. Figure~\ref{fig:r12L} displays this relation for dwarf galaxies
in the Local Group, including UFDs (grey squares). Data shown here are mostly taken from the continuously updated
\emph{Local Group and Nearby Dwarf Galaxies} database\footnote{\url{http://www.astro.uvic.ca/~alan/Nearby_Dwarf_Database.html}}
of \citet{McConnachie2012}. 
We supplement these data with recently discovered UFD candidates:
Eridanus IV \citep{cerny2021},
Bo\"otes V, Leo Minor I, Virgo II \citep{cerny2022,smith2022},
Tucana B \citep{sand2022}, Pegasus V \citep{collins2022},
Leo M, and Leo K \citep{mcquinn2023}.
A large fraction of UFDs ($L_\mathrm{V}<10^5\,L_{\odot}$) are very compact with $R_{1/2}$ smaller than $100\,\rm{pc,}$ with some extreme cases being more compact that $10\,\rm{pc}$.
Those observed galaxies are compared to predictions from a variety of numerical models shown by the coloured points
\citep{jeon2017,Revaz2018,applebaum2020,agertz2020,jeon2021a,prgomet2022,gutcke2022,sanati2023}.
While  simulations are in agreement with observations when above a luminosity of $10^5\,L_{\odot}$,
in the UFD regime, predictions diverge with several models, predicting a half-light radius of more
than one order of magnitude above the observations. Except for the \citet{jeon2021a} models, none are able to 
reproduce the compactness of observed UFDs. 
We note that in the \citet{jeon2021a} star formation scheme, 
approximately ten gas particles are converted into stellar particles. With the resolution used, this process 
results in a stellar Pop II cluster with a mass of $500\,\rm{M_\odot}$. Consequently, models below $10^4\,L_{\odot}$
are resolved with a maximum of 20 stellar particles (Myoungwon Jeon; private communication), which can make the 
derivation of the half-light radius difficult.

The origin of the over-prediction of the size of UFDs remains elusive.
To begin answering this question, it is important to decipher whether or not numerical issues could affect the sizes of simulated systems.
(i) A lack of resolution could naturally be at the origin of this discrepancy. 
However, this argument by itself is not sufficient, as the \citet{Wheeler_2019} model is the most resolved,
with a 30 solar mass resolution, while being well above the relation.
(ii) Spurious numerical heating is know to exist when collisionless particles with different masses are used for the stellar and dark matter components \citep{Revaz2018,ludlow2019,ludlow2021,ludlow2023}.
This heating leads to an artificial increase in the stellar component.
(iii) An over-estimation of the stellar feedback as well as a bursty star formation history 
could also lead to over-estimation of the size of the system.
Nevertheless, we note that over-estimation of the sizes of UFDs seems to be a general trend, 
despite the use of a variety of numerical schemes and  star formation and stellar feedback recipes.

Beyond numerical issues, one might ask whether or not the complex buildup history that any galaxy ---including
UFDs--- suffers during its formation in the hierarchical $\Lambda$CDM model
could be responsible for this tension.
Both observations and models agree that UFDs are dominated by very old stars. According to numerical models, 
star formation is inhibited after those tiny galaxies have lost their gas at the end of the EoR.
Consequently, being devoid of gas, UFDs evolve as collisionless systems during the remaining 
$12\,\rm{Gyr}$ or so up to the present time.
Without dissipation, any perturbation 
to the stellar component formed prior to the EoR will lead to an increase in its size.
In the following sections, we explore the extent to which the buildup of UFDs in a cosmological context can lead to 
oversized UFDs.

%

%%%%%%%%%%%%%%%%%%%%%%%%%%%%%%%%%%%%%%%%%%%%%%%%%%%%%%%%%%%%%%%%

\section{Methods and simulations}\label{sec:simulations}

%%%%%%%%%%%%%%%%%%%%%%%%%%%%%%%%%%%%%%%%%%%%%%%%%%%%%%%%%%%%%%%%
%

%fact that UFDs ceased to form stars at the end of the EoR,
%i.e, after about one Gyr after the Big Bang. In the following, we
%will thus assume that the stellar component is in place at z=6. 
%As the gas
%is evaporated by the UV-background heating, it is appropriate to consider that UFDs 
%will further evolve as a non-dissipative collisonless systems.

Here, we propose a new and numerically clean method
that takes advantage of the fact that UFDs lost their gas at the end of the EoR.
This fact allows us to assume that the stellar component of UFDs is in place at $z=6$ and
that UFDs further evolve as non-dissipative collisionless systems.
This allows us to study their buildup history and its impact on their size
using dark-matter-only (DMO) simulations.
%, neglecting the gas component from $z=6$ on.
%
%%We followed \citet{sanati2023} and tailored similar cosmological zoom-in simulations, however dark matter 
%%only (DMO) with an increase of 64 in the mass resolution, i.e., with dark matter particles with a mass of 
%%$77\,\rm{M_\odot}$.
We followed \citet{sanati2023} and tailored similar cosmological zoom-in simulations, but this time in a DMO regime
with an increased resolution.
In all simulations presented here, the dark matter particle mass is set to $77\,\rm{M_\odot}$, 
a factor 64 compared to \citet{sanati2023}.
Simulations were run with the code \textsc{Swift}\footnote{\url{https://swift.dur.ac.uk/}} \citep{schaller2016,schaller2018,schaller2023}, a modern gravity and smoothed-particle hydrodynamics solver for
astrophysical and cosmological applications that benefits from a task-based parallelism.
From $z=70$ down to $z=2.8$, we used a co-moving gravitational softening length of $35\,\rm{pc}$. 
This is further kept constant to $9.3\,\rm{pc}$ down to $z=0$ in physical coordinates.
We simulated a subset of 7 halos among the 18  
studied by \citet{sanati2023} that led to the formation of UFDs.
We choose the ones that have the faintest stellar component, for which the size discrepancy 
between observations and simulations is the greatest. 
Table~\ref{tab:ufds} shows the properties of those halos at $z=0$ as derived from \citet{sanati2023}: luminosity, stellar mass, virial mass, radius, half-light radius, and velocity dispersion.

To estimate the effect of the buildup history on the size of UFDs measured from the stellar component, which is absent from our DMO simulations, we used the following approach:
(i) extract UFDs at $z=0$,
(ii) find progenitor subhalos at $z=6$ (see Section~\ref{sec:subhalos} for a proper definition of subhalos),
(iii) determine the stellar content and properties  of each subhalo,
(iv) track the evolution of the stellar content, in particular its size, while assembling the UFDs, down to $z=0$.
Each step is detailed below.

\begin{table}[h]
  \caption{\small Global properties of the seven simulated  halos  taken from \citet{sanati2023}. }
  %abundance ratio of iron with respect to hydrogen. 
    \label{tab:ufds}
  \centering
  %\tiny
  \resizebox{0.49\textwidth}{!}{%
  \begin{tabular}{l c c c c c c  }
  \hline
  \hline

    Halo ID & $L_{\rm{V}}$           & $M_{\star}$            & $M_{\rm{200}}$          & $R_{\rm{200}}$  &  $R_{1/2}$ & $\sigma_{\rm{LOS}}$  \\
       & $[10^5\,\rm{L_\odot}]$ & $[10^5\,\rm{L_\odot}]$ & $[10^9\,\rm{M_\odot}]$  & $[\rm{kpc}]$  & $[\rm{kpc}]$  & $[\rm{km/s}]$  \\ 
  \hline
  \hline
  \texttt{h277} & 0.10 & 0.26 & 0.36 & 22.2 & 0.42 & 5.3  \\
  \texttt{h249} & 0.07 & 0.18 & 0.39 & 22.8 & 0.28 & 5.7  \\
  \texttt{h315} & 0.05 & 0.14 & 0.33 & 21.6 & 0.51 & 5.6  \\
  \texttt{h323} & 0.05 & 0.13 & 0.27 & 20.1 & 0.21 & 4.0  \\
  \texttt{h170} & 0.05 & 0.13 & 0.62 & 26.5 & 0.41 & 5.0  \\
  \texttt{h273} & 0.04 & 0.09 & 0.39 & 22.7 & 0.32 & 4.1  \\
  \texttt{h291} & 0.02 & 0.04 & 0.33 & 21.4 & 0.18 & 5.5  \\
  \hline
  \hline
  \end{tabular}%
  }

  \tablefoot{
  The first column gives the halo ID following \citet{Revaz2018}. $L_{\rm{V}}$ is the V-band luminosity and $M_{\star}$ the stellar mass obtained from the hydrodynamical simulations.
  $M_{\rm{200}}$ is the virial mass, i.e. the mass inside the virial radius $R_{\rm{200}}$. $R_{1/2}$ is the half-light radius, and $\sigma_{\rm{LOS}}$ is the line-of-sight velocity dispersion of the stellar component. 
  }

\end{table}

\subsection{Main halo extraction}

For each DMO zoom-in simulation, the main halo, that is, the one that hosts the UFDs at $z=0,$ is found 
using the position of the corresponding halo from the lower-resolution hydrodynamics runs of \citet{sanati2023}.
All dark matter particles inside the virial radius are extracted. 

\subsection{Subhalo extraction}
\label{sec:subhalos}

Relying on their IDs, those main halo DM particles are extracted from the entire cosmological box at $z=6$. 
As the final main halo is not in place  at this redshift (see Sect.~\ref{sec:halo_prop_z6}), these particles 
define a subset of halos; we call them DMO {subhalos}. We define subhalos as clusters of particles for 
which the virial radius can be determined, that is, concentration of matter with a mean density larger than 200 times the critical density. Particles belonging to the main halo but not part of subhalos are ignored.

\subsection{Stellar mass determination}

We take advantage of the hydrodynamics runs to determine the
stellar content of our DMO subhalos.
Firstly, following the exact same method as the one used for the DMO runs\footnote{Here, the virial mass 
is computed including the dark matter but also baryons.}, we extracted subhalos at $z=6$ from
the hydrodynamics runs. 
Some of those halos ---usually the most massive ones--- contain stellar particles that will contribute to the 
formation of the stellar content of the UFD at $z=0$. 
For each subhalo $i$, we record the number of stellar particles it contains,  $N_{\star,i,\rm{hydro}}$, whereby completely dark halos have $N_{\star,i,\rm{hydro}}=0$.
%we cross-match our DMO subhalos with the ones existing in the hydro-run
%for which the stellar content, i.e. the number of stellar particles  $N_{\star,i}$ is known, where
%$i$ refers to a given subhalo.
Secondly, we cross-match the DMO and hydrodynamics subhalos to find corresponding halo pairs.
As the two simulations differ in their resolution and baryonic content, 
both the position and the mass of halos can vary slightly \citep{garrison-kimmel2017}, necessitating some care in the identification of those pairs.
Knowing the position of all halos as well as their mass, we identified corresponding 
halos by minimising a norm in configuration space and checking that their halo mass corresponds.
A final visual check was carried out to guaranty the perfect match.
We find that, on average, halos are slightly globally shifted between the two runs.
%This provides each DMO subhaloe $i$ with a stellar mass content $M_{\star,i}$.
We then attribute an associated number of stellar particles $N_{\star,i}$  to each DMO subhalo using a conversion factor $f_{\star}$, such that $N_{\star,i} =  {\rm int}\left(  f_{\star}\cdot N_{\star,i,\rm{hydro}} \right)$.
Finally, for each subhalo $i$, we find the $N_{\star,i}$ most gravitationally bound DM particles, which we further 
considered as stellar particles, conserving their properties at the time of selection, that is, their position, 
velocity, and mass. 
This is done by 
(i) extracting only particles within the virial radius of each subhalo and
(ii) computing their total energy with respect to the subhalo.
%To get the bounded energy, we need to compute both the kinetic and potential energy.
%The former is obtained by substracting to each particle velocity the mean velocity $\vec{\bar{v}}_{i}$ of each halo $i$.
%The second is obtained by a tree-code, computing the potential $\Phi_{j,i}$ energy of each particles $j$ in $i$.
The energy $E_{j,i}$ of each particle $j$ is written as: 
\begin{equation}
  E_{j,i} = \frac{1}{2} \left(  \vec{v}_{j} - \vec{\bar{v}}_{i}  \right)^2 + \Phi_{j,i},
\end{equation}
where $\vec{\bar{v}}_{i}$ is the mean velocity of the halo $i$ and $\Phi_{j,i}$ is the potential
energy of the particle $j$ considering only the mass in the virial radius.
The $N_{\star,i}$ most gravitationally bound particles are then the ones with minimal energy.
These are assumed to represent the stellar content of the subhalo and are referred to here as {a cluster}. 
We note that through this procedure, we maximise the compactness and robustness with respect to gravitational perturbations of the stellar cluster defined. This is an important
choice, and is further discussed below.

The value of the factor $f_{\star}$ can be modified to vary the number of particles representing a cluster
while keeping the same ratio between them, that is, the stellar content as determined by the hydro-runs.
Its fiducial value $f_{\star,\rm{fid}}$ is set to $32$,  that is, we define 32 times more particles compared to the 
number of stellar particles\footnote{We ignored stellar particles representing metal-free (population III) stars as those are assumed to be short lived and will not contribute to the UFDs stellar population at $z=0$.} found in 
a subhalo in the lower-resolution hydrodynamics run. This number is motivated by the current resolution, which leads an 
increase of 64 (in term of number of particles) compared to the \citet{sanati2023} simulations. 
However, it must be borne in mind
that those particles do not precisely represent stars but  are used as mass tracers.
As we show below, our fiducial choice also guarantees a sufficiently large number of tracers, which is essential
in order to properly define cluster size and obtain statistically significant results.
The impact of varying $f_{\star}$ is studied in Section~\ref{sec:fstar_influence}.
We also note that, as cluster particles are initially DM particles, their number,
within each cluster, scales with the cluster mass.

\subsection{Tracking the evolution of clusters}

Once clusters have been defined at $z=6$ as the most gravitationally bound DM particles, we can easily trace their
properties throughout the evolution of the formation of the UFDs. To this purpose, they are extracted
from the 878 snapshots saved between $z=6$ and $z=0$.
We  computed their mass and velocity centres as well as their 2D half mass.
As UFDs are composed exclusively of old stars \citep{brown2012}, the mass-to-light ratio is
constant and the half mass can replace the half light to a very high level of accuracy. 
This half mass is denoted $R_{1/2}$. To accurately determine its value, we used two methods.
In the first, we took advantage of the Plummer model property for which the scale radius
parameter is precisely the 2D half-mass radius. 
We then computed the surface density of the cluster assuming cylindrical symmetry and
fit it with a Plummer model to determine the scale radius that provides
a good estimation of $R_{1/2}$.
In the second method, we computed the cumulative mass profile through a logarithmic grid, 
also assuming cylindrical symmetry, and determined the radius where the value is half of the maximum.
In both methods, in order to avoid bias owing to a specific choice of line-of-sight,
we averaged the result obtained using seven random lines of sight.
As no major differences were found between the two methods, we used the second one.

We finally emphasise here that, by using DMO simulations, all particles share the same mass.
This avoids the aforementioned numerical bias where stellar particles with smaller
masses are numerically heated up by more massive dark matter ones 
\citep{Revaz2018,ludlow2019,ludlow2021,ludlow2023}; this is an essential factor in providing reliable 
cluster sizes.

%
%%%%%%%%%%%%%%%%%%%%%%%%%%%%%%%%%%%%%%%%%%%%%%%%%%%%%%%%%%%%%%%%

\section{Results}\label{sec:results}

%%%%%%%%%%%%%%%%%%%%%%%%%%%%%%%%%%%%%%%%%%%%%%%%%%%%%%%%%%%%%%%%

\begin{figure*}
  %\vspace{-20pt}

  \begin{subfigure}[b]{1\textwidth}         
  \includegraphics[width=0.49\textwidth]{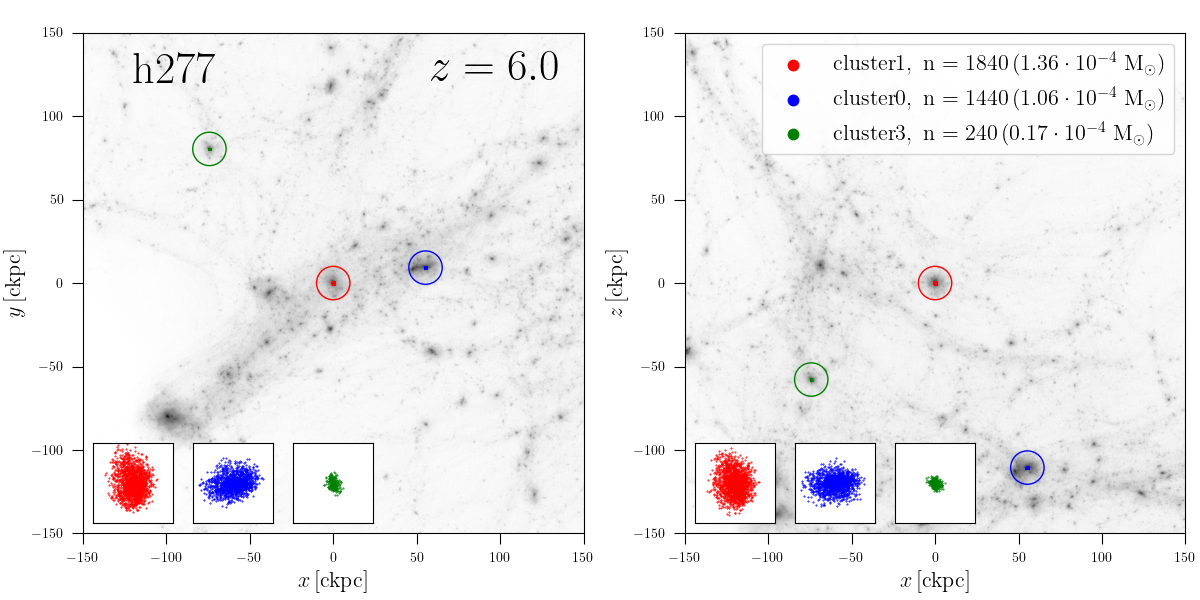}
  \includegraphics[width=0.49\textwidth]{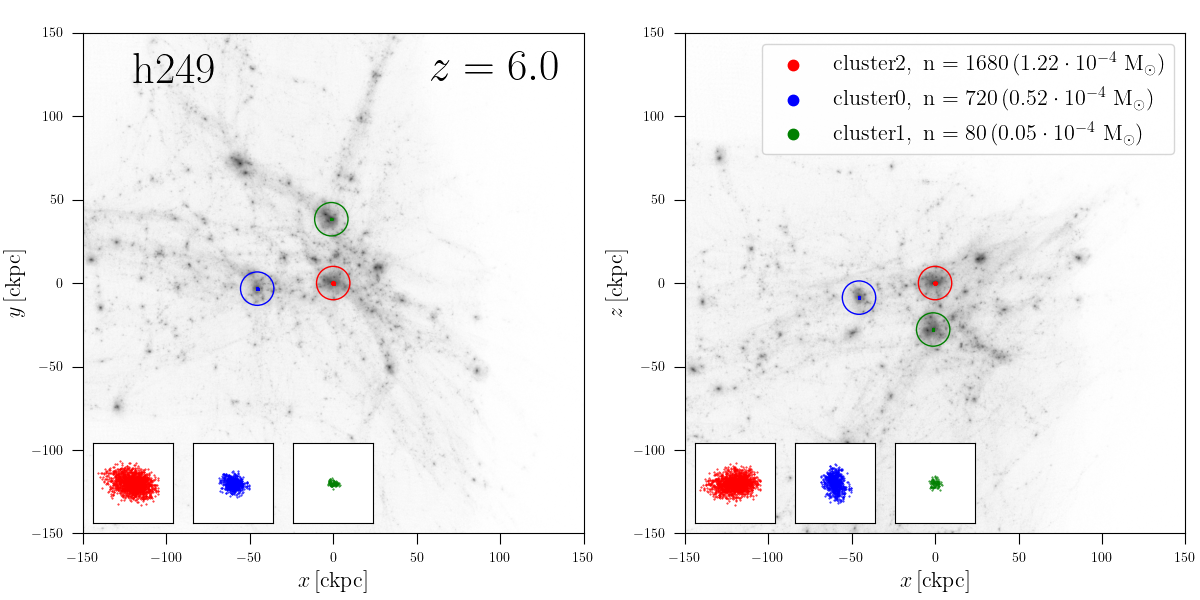}
  \includegraphics[width=0.49\textwidth]{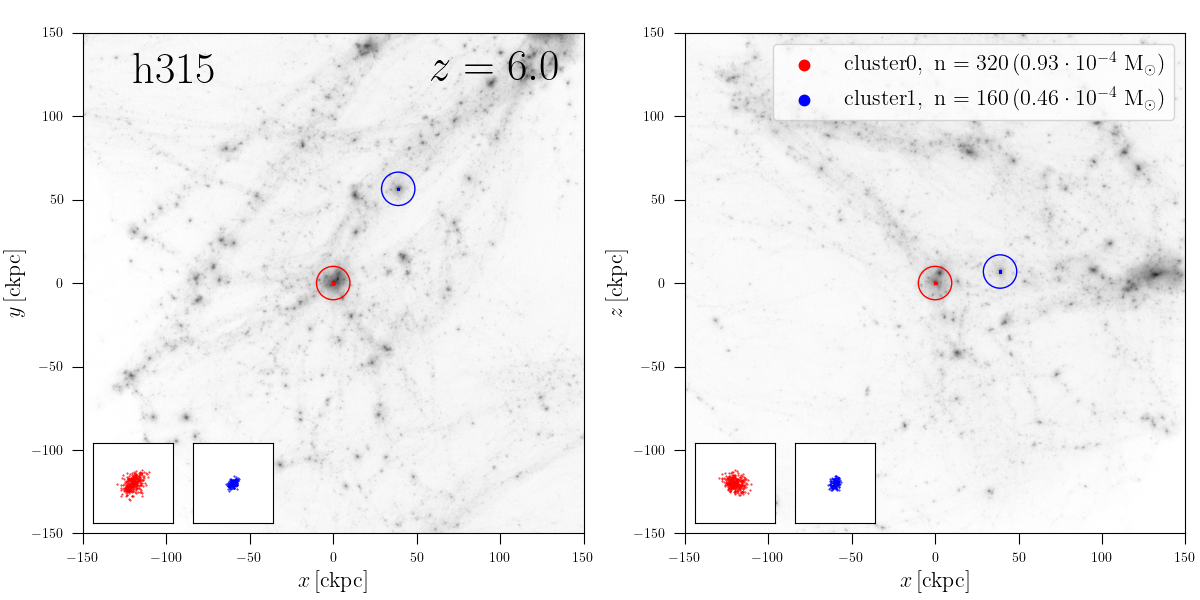}
  \includegraphics[width=0.49\textwidth]{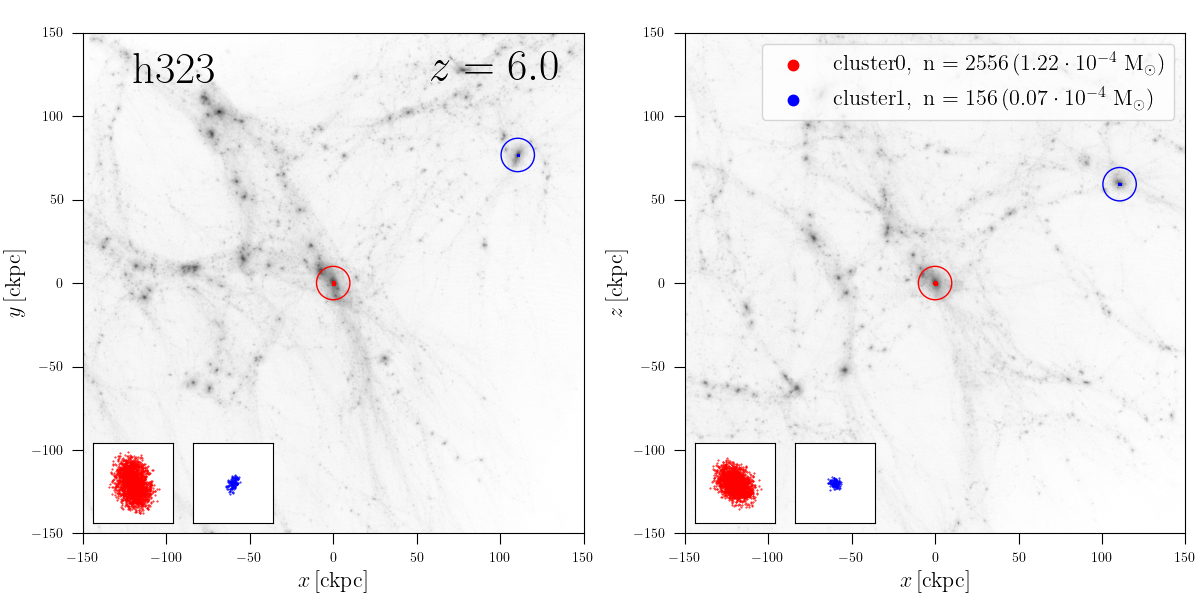}
  \includegraphics[width=0.49\textwidth]{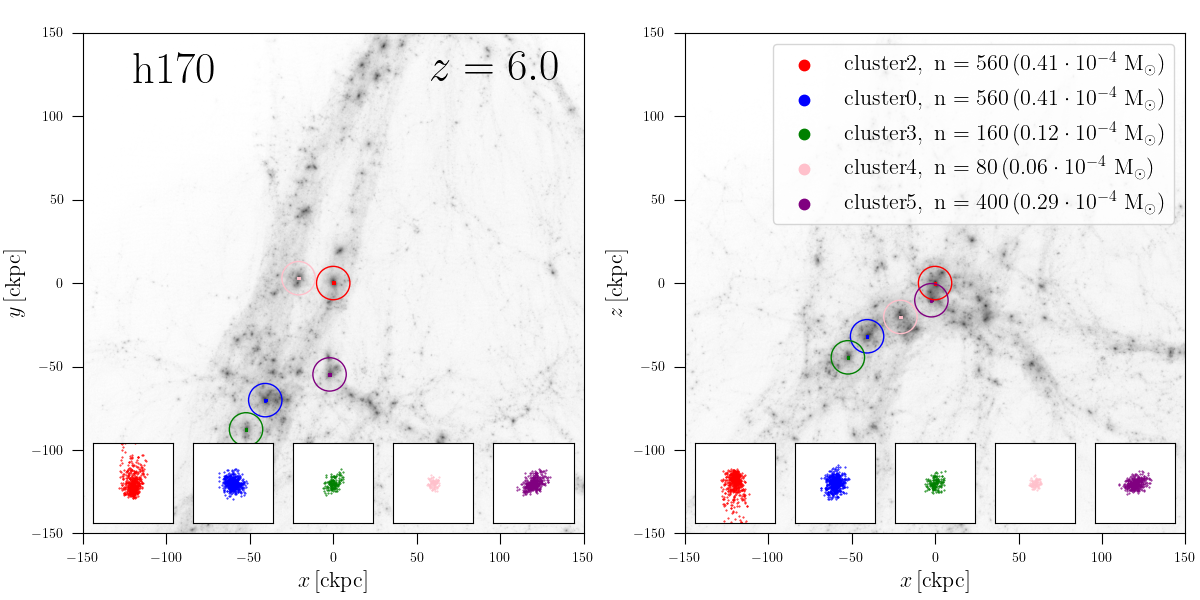}
  \includegraphics[width=0.49\textwidth]{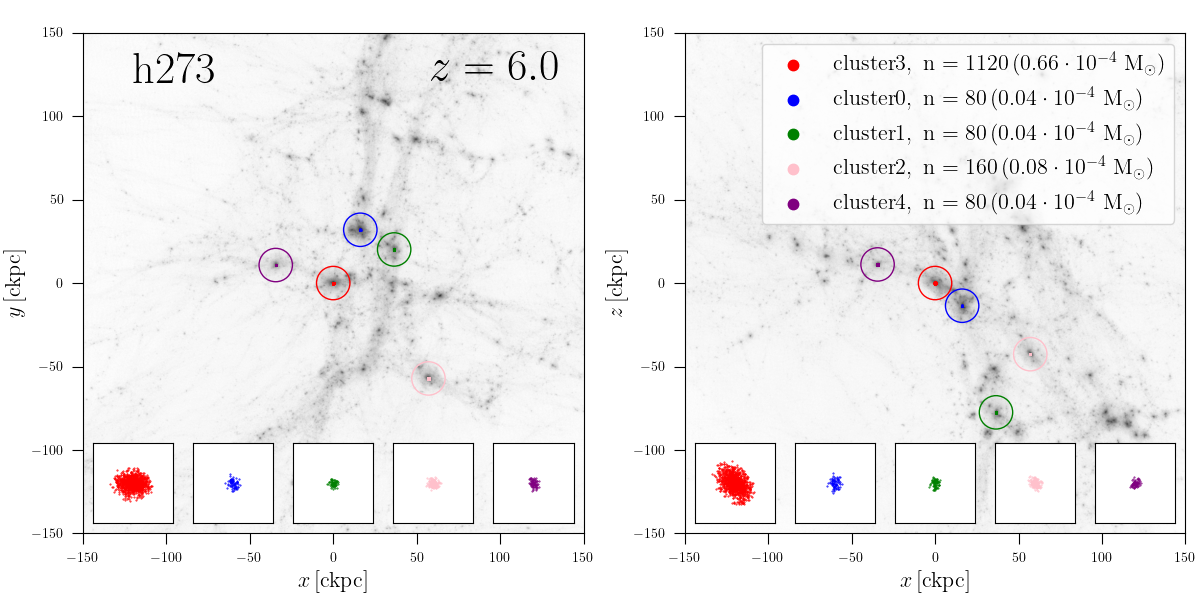}
  \includegraphics[width=0.49\textwidth]{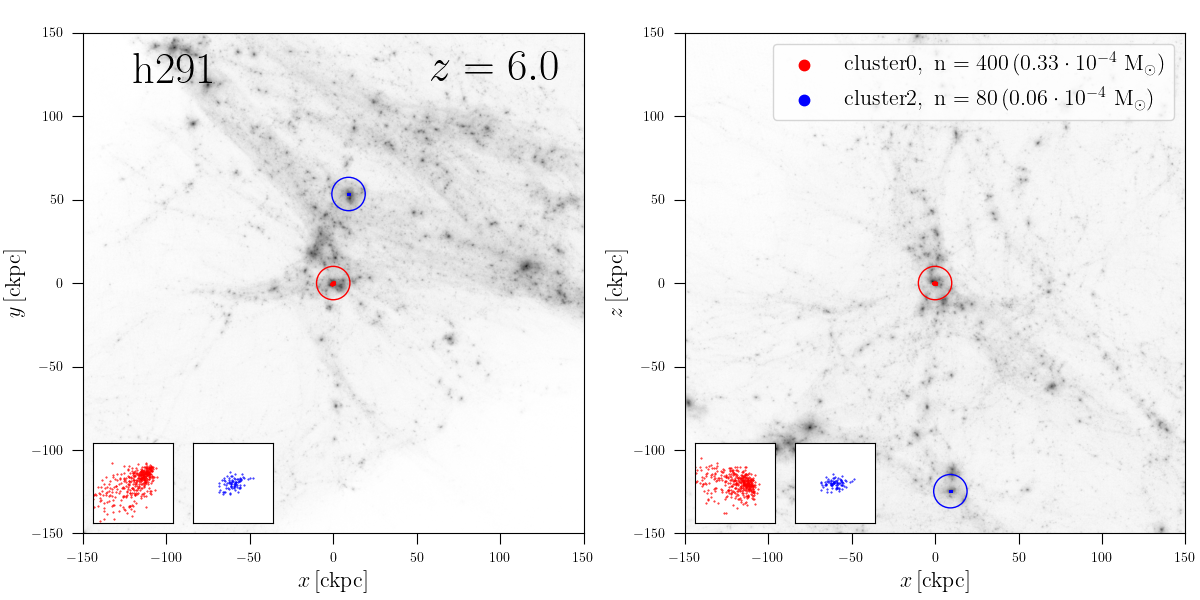}
  \end{subfigure}
  
\caption{Relative positions of the selected clusters at $z=6$ for each of the seven UFDs simulated. Clusters are 
indicated with a specific colour on top of the DM surface density in grey.
Red ones correspond to the clusters containing the largest numbers of particles. 
The subpanels at the bottom show the projection in proper coordinates ($100\,\rm{pc}\times 100\,\rm{pc}$) of each cluster. For each UFD, we show two different projections.}

\label{fig:mapz6}

\end{figure*}

%%%%%%%%%%%%%%%%%%%%%%%%%%%%%%%%%%%%%%%%%%%%%%%%%%%%%%%%%%%%%%%%
\subsection{Subhalo properties at z=6}
\label{sec:halo_prop_z6}

% goal : show that cluster are physically distinct

%For each halo, show the halo physical distribution
%in configuration and velocity space.
%Discuss the relative kinetic energy.
%(assume the subhalos to be decoupled from the Hubble flow)

Figure~\ref{fig:mapz6} shows the relative position of the 
selected clusters at $z=6$ for each of the seven simulated UFDs. 
Each cluster is displayed with a specific colour and over-plotted on top
of the surface density of the DM. 
All figures are centred on the mass-dominant cluster, which is  the one that contains the largest number
of particles and therefore the most mass, and is marked in red.
The list of clusters is provided in the  upper right legend, together with
the number of particles it contains.
This number reflects the stellar particle mass found in the hydrodynamics 
run; its absolute value depends on the parameter $f_{\star}$.

It is striking from this figure that, at $z=6$, which is the end of the EoR and also corresponds
to the end of the star formation history of UFDs, none of the seven studied systems is in place.
On the contrary, clusters that
will form a dwarf at $z=0$ are spread out and are fully dynamically disconnected, with 
relative distances that extend up to $130\,\rm{ckpc}$ (\texttt{h291}).
We recall that from this redshift down to $z=0$, those stellar clusters
will not dissipate energy except through dynamical friction. 
This underlines the difficulty those clusters ---spread out in physically distinct  mini-halos---
may face in order to end up forming very compact systems, such as the observed UFDs.

In the seven simulated halos, the number of identified clusters ranges from two (\texttt{h315}, \texttt{h291}) to five (\texttt{h170}, \texttt{h273}) (see Table~\ref{tab:ufds2}).
This number correlates with the virial mass at $z=0$. More massive halos, such as \texttt{h170} ($M_{200}=8.3\cdot 10^8\,\rm{M_\odot}$), are made up of five clusters, while
low-mass halos (\texttt{h323}, \texttt{h315}, $M_{200}<4.5\cdot 10^8\,\rm{M_\odot}$) have only two. This naturally reflects the hierarchical buildup
of the $\Lambda$CDM paradigm. The probability that more massive halos form from more massive subhalos
that contain stars is larger than for less massive ones.
We note that following our choice of $f_{\star,\rm{fid}}$, each cluster contains a relatively large number of particles ---that is, from 40
to 2556---, which is essential in order to obtain reliable and accurate properties during their evolution.
The shape of the extracted clusters is shown at the bottom of each panel of Fig.~\ref{fig:mapz6},
where particles  are projected in a $100\,\rm{pc}  \times  100\,\rm{pc}$ box in proper coordinates.
The size of each cluster ($R_{1/2,\rm{init}}$) 
is given in Table~\ref{tab:ufds2}.
We note that except for model \texttt{h291}, which is discussed in Sect.~\ref{sec:complex_environement}, 
they are all relatively spherical and compact, with a half mass of less than or roughly equal 
to $25-30\,\rm{pc}$. 
We find all clusters to be isotropic in velocity space with rather low 
line-of-sight velocity dispersions, ranging between $0.95$ and $2.2\,\rm{km/s}$.
Those cluster properties validate our selection method.

%{\bf With relative kinetic energy existing between different stellar clusters leading to the formation of the final UFD,
%and without the possibility to loose this energy (except through dynamical friction), 
%it is hard to understand how the stellar components initially spread out in physically distinct  mini-halos can 
%merge and form a final stellar system as compact as observed, with $R_{1/2} < 30\,\rm{pc}$ in the most extreme cases.}

%%%%%%%%%%%%%%%%%%%%%%%%%%%%%%%%%%%%%%%%%%%%%%%%%%%%%%%%%%%%%%%%
\subsection{The buildup}

The left panels of Figs.~\ref{fig:R12vsRedshift1} and \ref{fig:R12vsRedshift2} show the time evolution of cluster size traced by their half-mass radius, shown as a function of redshift.
The red curves correspond to the mass-dominant clusters. 
%
%Their initial size $R_{1/2,\rm{init}}$ at $z=6$
%is given in Tab.~\ref{tab:ufds2} together with the final UFDs size $R_{1/2,\rm{fin}}$, 
%drawn as the black horizontal curve.
%We first notice that except for model \texttt{h291} that will be discussed below, 
%the half-mass of all dominant clusters are pretty compact, less or roughly equal 
%to $20\,\rm{pc}$. This validates our selection method.
%
The sizes of the formed UFDs at $z=0$ ($R_{1/2,\rm{fin}}$) are given in Table~\ref{tab:ufds2},
and are drawn as the black horizontal curve.
All clusters have a size above the resolution of the gravity force, which is shown 
by the dotted lines.
This guarantees safe numerical treatment and prevents uncontrolled heating by two-body
relaxation. 
With our choice of $f_{\star,\rm{fid}}$, which favours a larger number of particles in clusters,
and consequently a larger initial size, no cluster is initially found with a half mass of less than $10\,\rm{pc}$.
 In Sect.~\ref{sec:fstar_influence}, we show the influence of decreasing the number of particles
in clusters  on their size.
Figure~\ref{fig:MhalovsTime} complements Figs.~\ref{fig:R12vsRedshift1} and \ref{fig:R12vsRedshift2} by
showing the time evolution of the virial mass of the halo that hosts the main cluster and corresponds to the
halo of the final UFD.

%{\bf 20 pc is already bigger than some UFDs... No cluster is found with 
%with a radius < thatn 10pc. This is a problem for a fair comparizon with UFDs.
%shall we relate that to the fact the at z=6, the environement
%is complext ? No clean subhalo ? Or is it a problem of resolution ? This can be adressed with fstar}

%
\begin{table}[h]
  \caption{\small Properties of the seven halos in the DMO run. $R_{1/2,\rm{init}}$ is the half-mass radius
  of the dominant cluster at $z=6$.}
  %abundance ratio of iron with respect to hydrogen. 
    \label{tab:ufds2}
  \centering
  %\tiny
  \begin{tabular}{l c c c c c c c }
  \hline
  \hline

    Halo ID & 
    $M_{\rm{200}}$ &
    $R_{1/2,\rm{init}}$     &  
    $R_{1/2,\rm{fin}}$    &
    $\sigma_{\rm{LOS,fin}}$   &
    \#  \\

    &
    $[10^9\,\rm{M_\odot}]$ &
    $[\rm{pc}]$            & 
    $[\rm{pc}]$            &
    $[\rm{km/s}]$          &
    \\ 
  \hline
  \hline
  \texttt{h277} & 0.49 & 28.99 & 96.86  & 4.02 & 3\\
  \texttt{h249} & 0.53 & 21.86 & 59.14  & 3.73 & 3\\
  \texttt{h315} & 0.42 & 14.21 & 53.48  & 3.03 & 2\\
  \texttt{h323} & 0.38 & 25.80 & 36.08  & 2.96 & 2\\
  \texttt{h170} & 0.83 & 20.29 & 89.34  & 4.41 & 5 \\
  \texttt{h273} & 0.49 & 20.05 & 48.33  & 3.40 & 5\\
  \texttt{h291} & 0.49 & 34.82 & 77.82  & 3.40 & 2\\
  \hline
  \hline
  \end{tabular}%

  \tablefoot{
  $R_{1/2,\rm{fin}}$ is the half-mass radius of the UFDs at $z=0$, obtained
  through the contribution of all clusters. $M_{\rm{200}}$ is the virial mass.
  We note that it is systematically larger compared to the value derived from the hydro-runs (see Table~\ref{tab:ufds}).
  This is due to the fact that in the latter simulations, the gas has been lost at the end of the EoR.
  $\sigma_{\rm{LOS,fin}}$ is the line-of-sight velocities of clusters at $z=0$. The last column gives the
  number of clusters found in each halo.  
  }

\end{table}

\subsubsection{Complex high-redshift environment}
\label{sec:complex_environement}

The large initial size of cluster \texttt{h291} (however with a lower number of particles
compared to others) is related to its complex environment. Between redshifts
$7$ and $5$, the subhalo that hosts the majority of the stars in the hydrodynamics run is not fully in place and 
is permanently disturbed by the accretion of slightly smaller halos. Therefore, finding a compact and
strongly gravitationally bound cluster in this halo is a difficult task. 
While its value at $z=6$ is $35\,\rm{pc}$, Fig.~\ref{fig:R12vsRedshift2} shows strong variations
with time, which is evidence of the hostility of the environment.
At $z=4$, the size of this dominant cluster reaches $75\,\rm{pc}$, a value that is kept constant
until the cluster merges with a second cluster at $z=0.7$ leading to a final size of $79\,\rm{pc}$.

\subsubsection{Nearly isolated subhalos}

\begin{figure}
\includegraphics[width=0.49\textwidth]{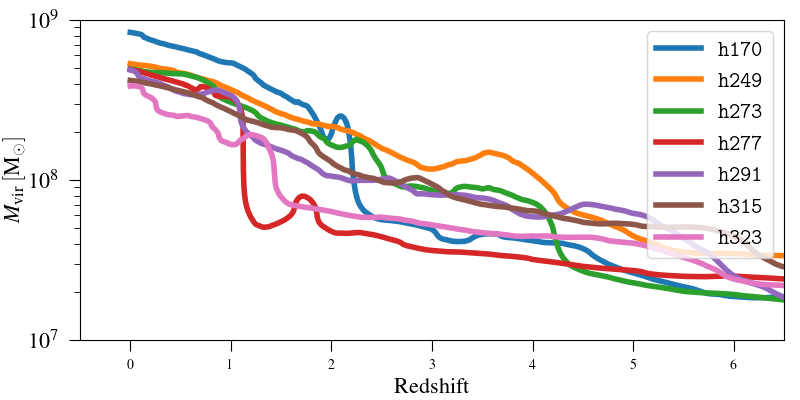}
\caption{Evolution of the virial mass of the dominant halo (i.e. the halo hosting the dominant cluster) as a function of redshift for each simulated UFD.
}
\label{fig:MhalovsTime}
\end{figure}

\texttt{h315} and \texttt{h323} are two much simpler cases, each
involving only two clusters. 
Starting from about $14\,\rm{pc}$, the size of the mass-dominant cluster 
(\texttt{cluster0}) of \texttt{h315} does not increase 
significantly until $z=3,$ when it suddenly suffers the gravitational influence
of two dark halos (at respectively  $z\cong 2$ and $z\cong 1$). This is sufficient to  
puff-up its size up to $25\,\rm{pc}$ at $z=1$. 
The merger with the second cluster (\texttt{cluster1}) is the major factor responsible for its final size of
$R_{1/2,\rm{fin}}=56.9\,\rm{pc}$. The main halo of \texttt{h323} does not suffer any 
strong perturbation until $z=1.3,$ where it suddenly interacts with a star-free dark halo 
that merges at $z=1$. 
This merger induces an abrupt increase in the halo virial mass as seen in Fig.~\ref{fig:MhalovsTime}.
We note that in this DMO run, the second cluster (\texttt{cluster2}) 
has still not merged at the end of the simulation, contrary to the hydrodynamics run. This is 
not surprising given the differences between the two runs in terms of
resolution and baryonic physics, which leads to differences in the dark halo 
masses \citep[see e.g.][]{garrison-kimmel2017}. 
It is clear that if this second cluster had merged, it would have contributed
to a further increase in the size of the final UFD.
With virial masses of $4.2\cdot 10^{8}$ 
and $3.8\cdot 10^{8}\,\rm{M_\odot}$ (see Table~\ref{tab:ufds2}) respectively, \texttt{h315} and 
\texttt{h323} are the lightest halos among our sample. This explains their 
simple buildup history, which results in relatively compact UFDs, albeit less compact 
than some observed UFD candidates.

\subsubsection{Complex buildup histories}

Increasing the halo mass leads to more complex histories with more
clusters involved.
In \texttt{h277} and \texttt{h249}, three clusters participate in the
formation of a UFD, and in
\texttt{h170} and \texttt{h273},  five clusters participate in each.
In these four cases, the impact of cluster mergers is extremely clear, as seen from the evolution
of the cluster size.
In extreme cases, such as \texttt{cluster0} in \texttt{h249}, $R_{1/2}$ jumps from $20$ up to $300\,\rm{pc}$.
Overall, these mergers lead to a strong expansion of the stellar system, with a final half-mass
radius above $45\,\rm{pc}$ (see Table~\ref{tab:ufds2});  the case of \texttt{h170} is up to $100\,\rm{pc}$.
This corresponds to an increase 
in the initial mass-dominant
cluster size by a factor of between $2.5$ and 5.

\begin{figure*}
  %\vspace{-20pt}

  \begin{subfigure}[b]{1\textwidth}         
  \includegraphics[width=0.39\textwidth]{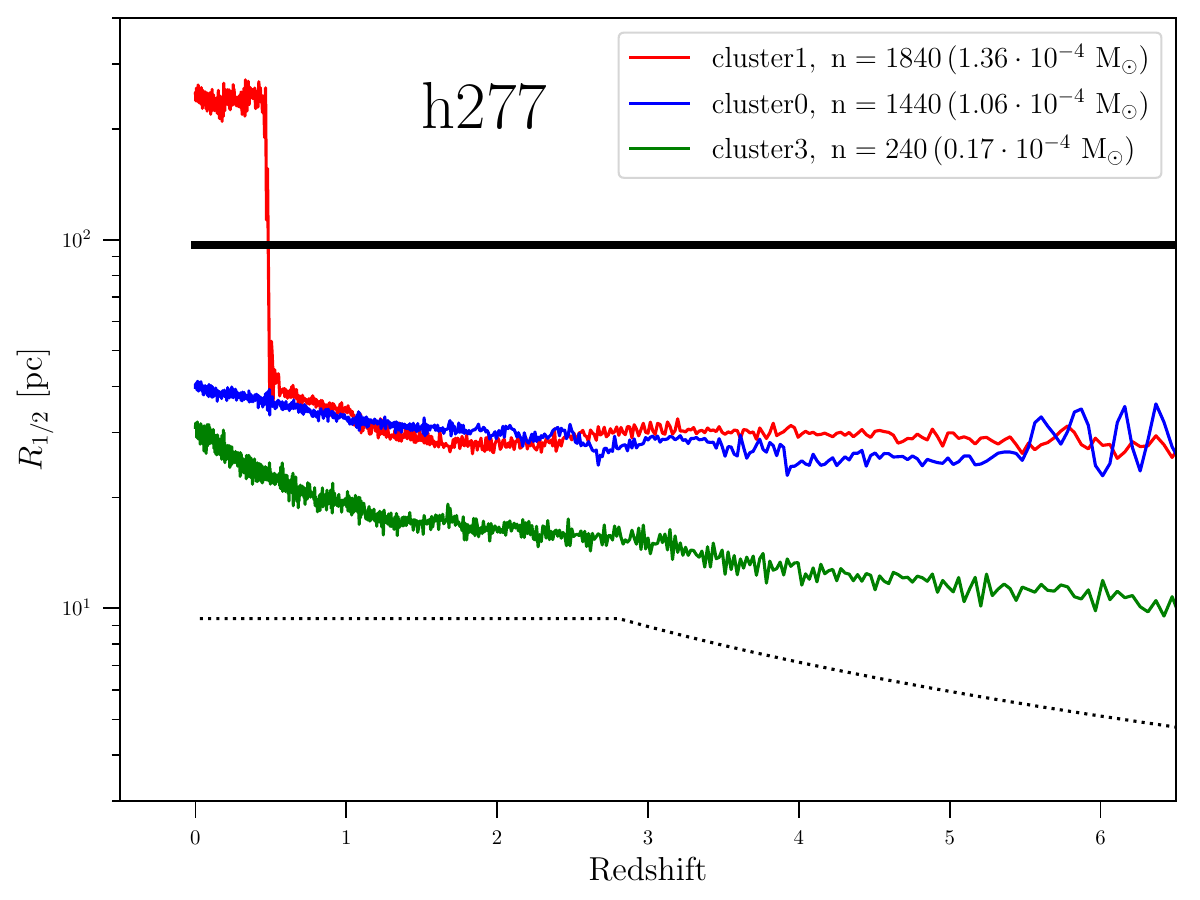}\includegraphics[width=0.60\textwidth]{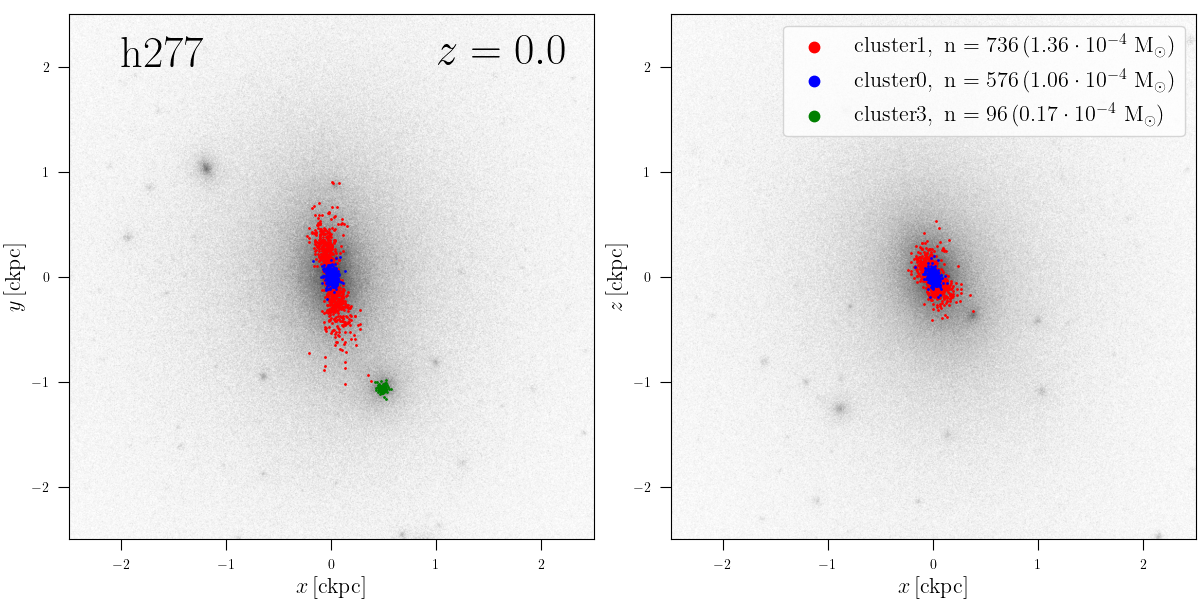}
  \includegraphics[width=0.39\textwidth]{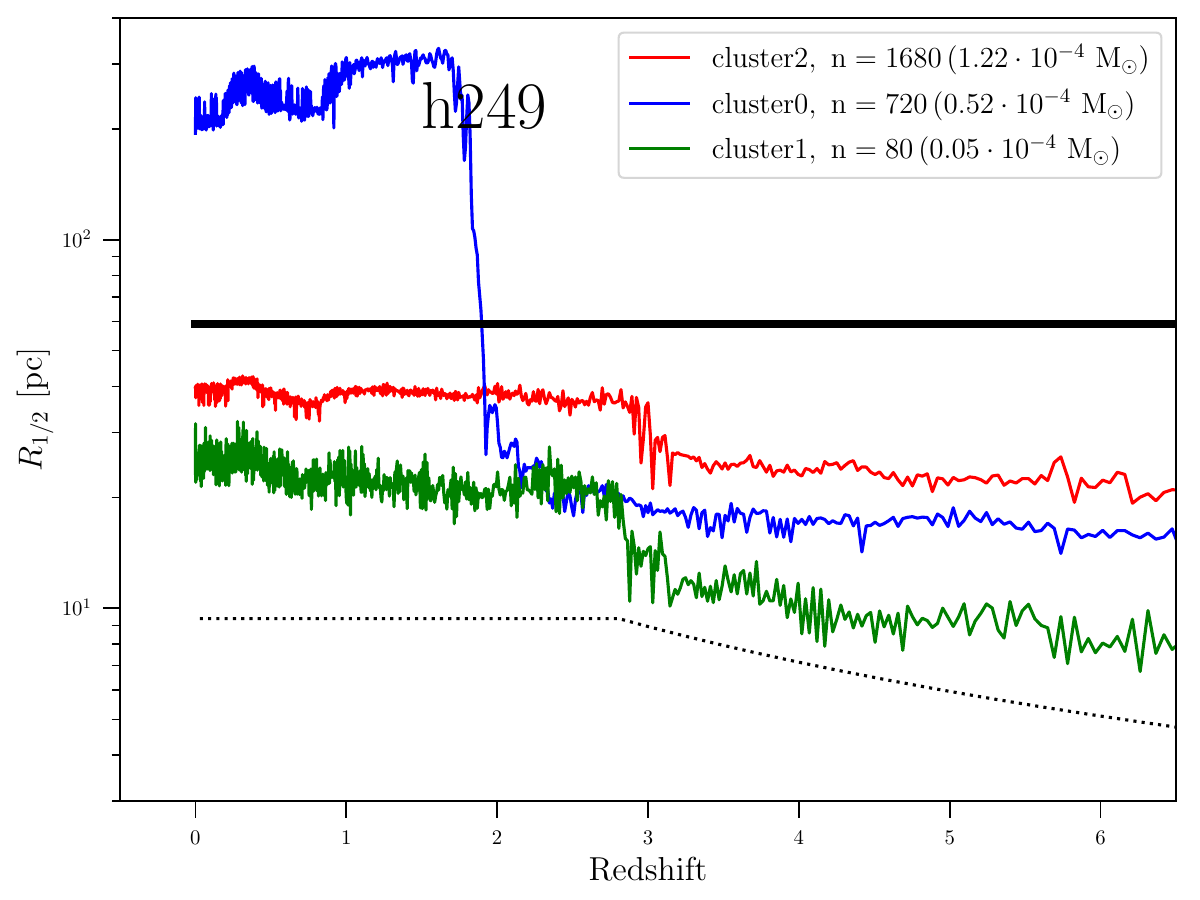}\includegraphics[width=0.60\textwidth]{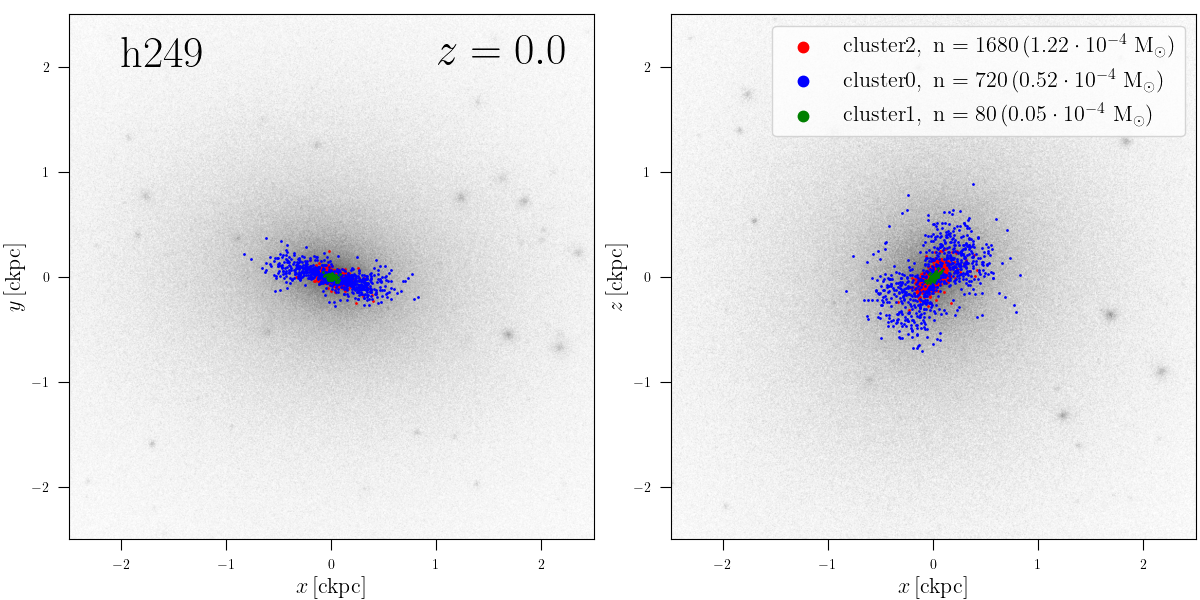}
  \includegraphics[width=0.39\textwidth]{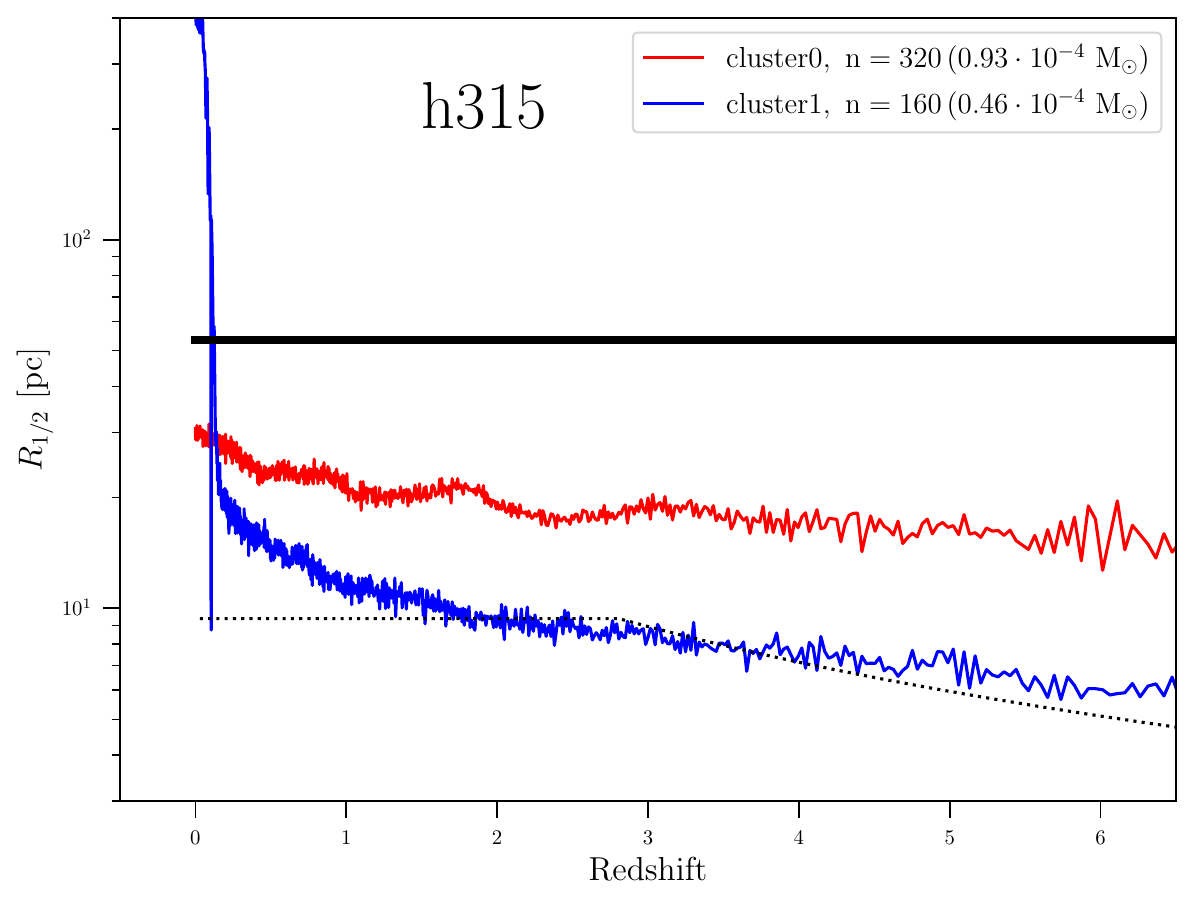}\includegraphics[width=0.60\textwidth]{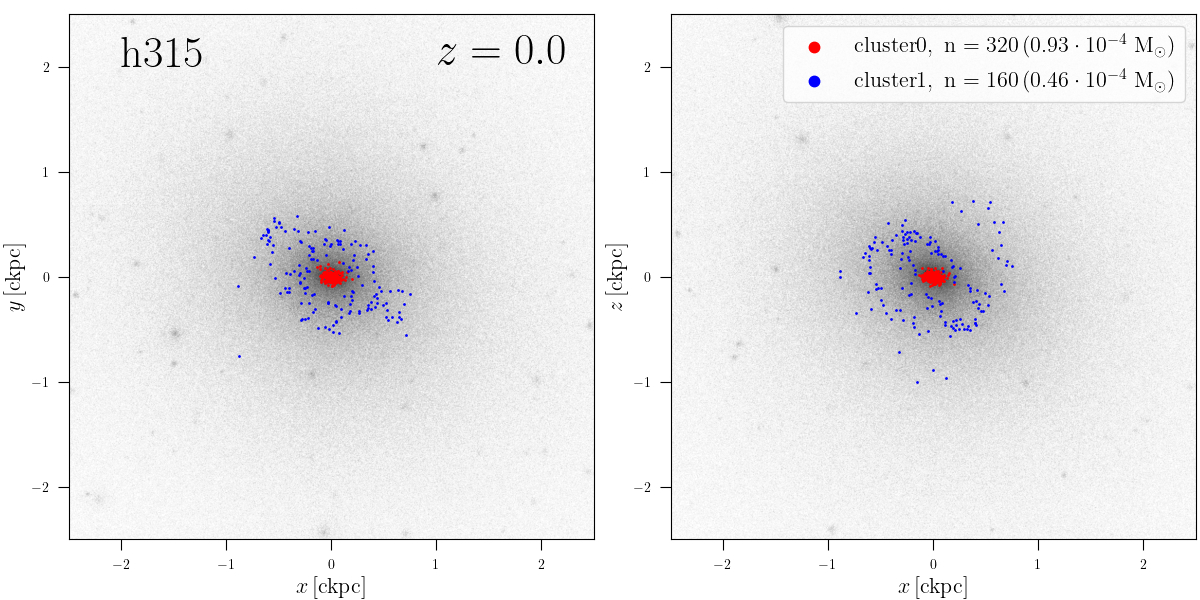}
  \includegraphics[width=0.39\textwidth]{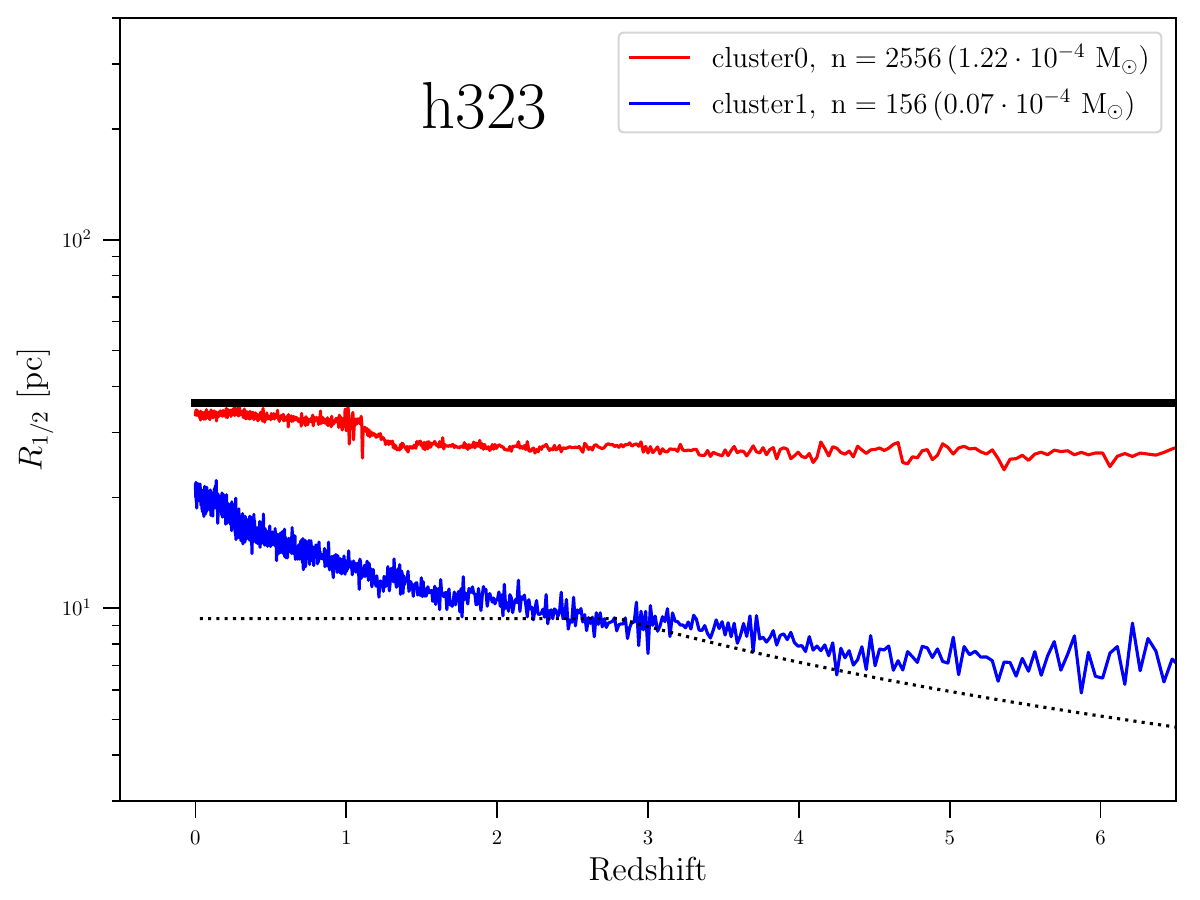}\includegraphics[width=0.60\textwidth]{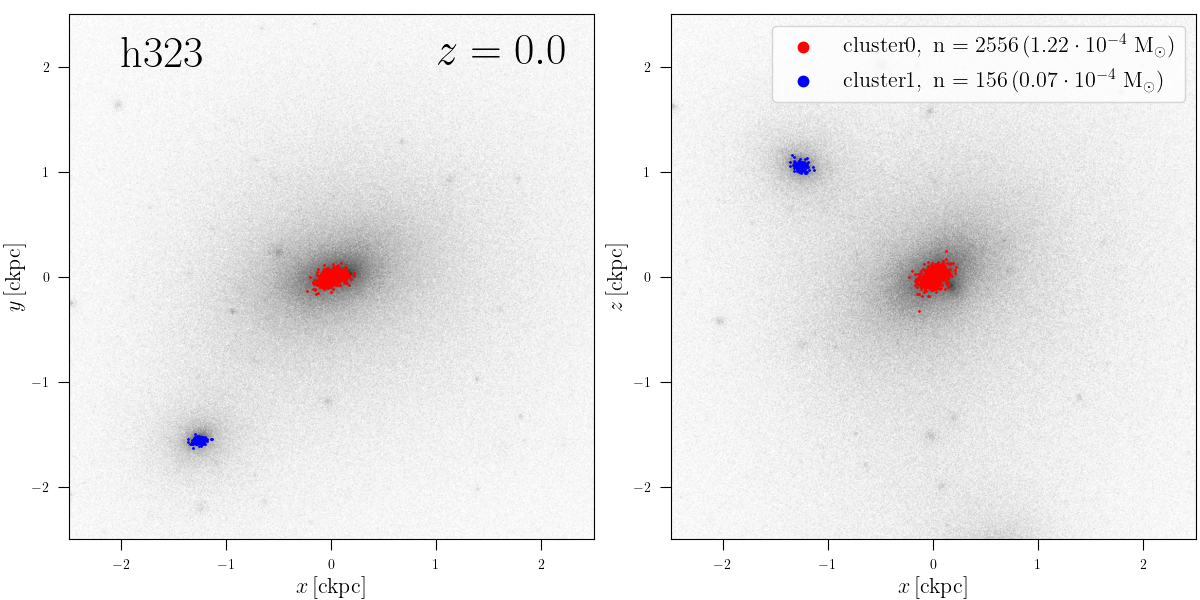}
  \end{subfigure}
  
  \caption{
  The buildup history of UFDs and their final shape.
  \emph{left}: Evolution of cluster size in proper units for the four models 
  \texttt{h277}, \texttt{h249}, \texttt{h315,} and \texttt{h323}. 
  Each line corresponds to one cluster and follows the colour code of Fig.~\ref{fig:mapz6}, the red one being
  the main cluster, defined as  the one with the largest number of particles.
  The dotted line at the bottom of the plot indicates the resolution of the simulation, i.e. the gravitational
  softening length. The horizontal thick line corresponds to the final size of the entire stellar system, i.e. 
  considering all clusters together. 
  \emph{right}: Two projections of the final stellar distribution of the UFDs  at $z=0$.  Particles are coloured according to their membership to clusters. The grey map in the  background corresponds to the DM surface density.
  }
  
  \label{fig:R12vsRedshift1}

\end{figure*}

\begin{figure*}
  %\vspace{-20pt}

  \begin{subfigure}[b]{1\textwidth}         
  \includegraphics[width=0.39\textwidth]{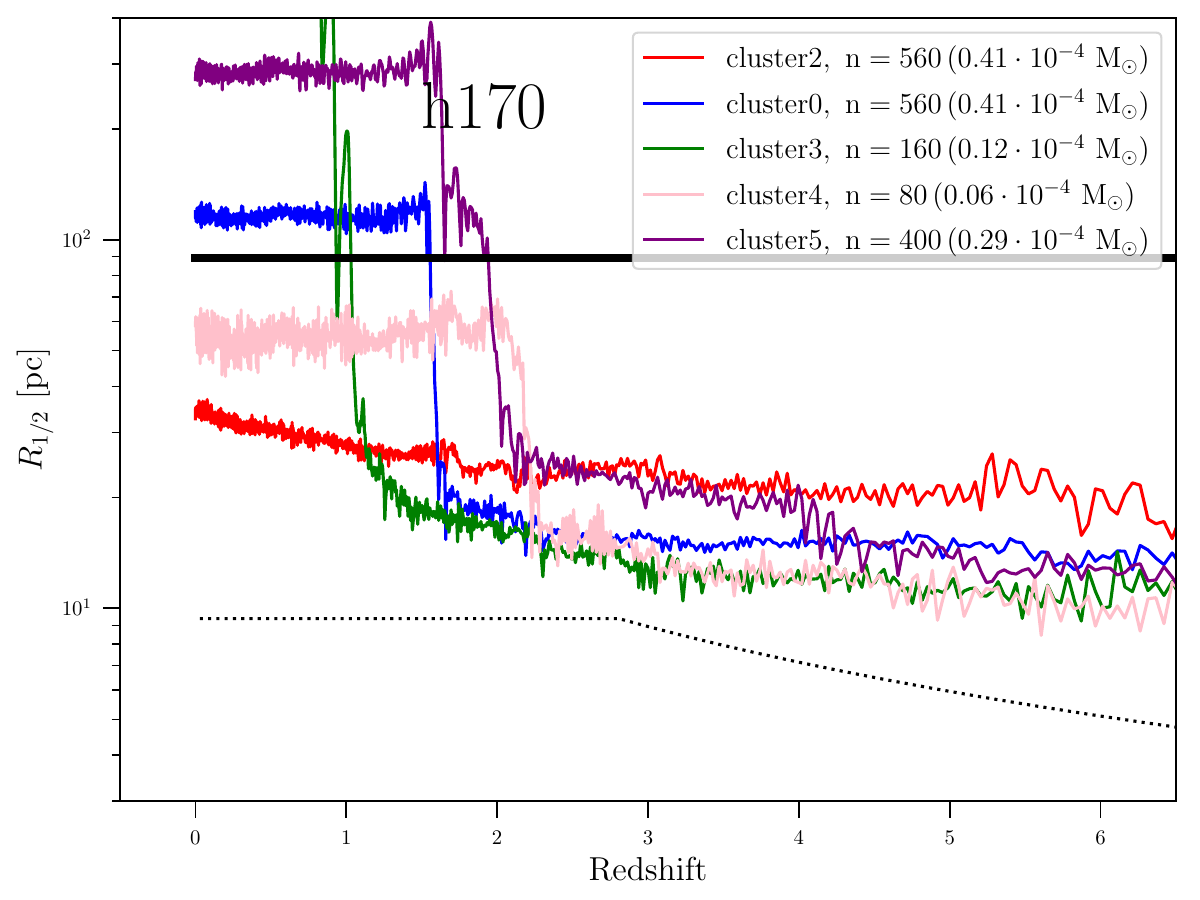}\includegraphics[width=0.60\textwidth]{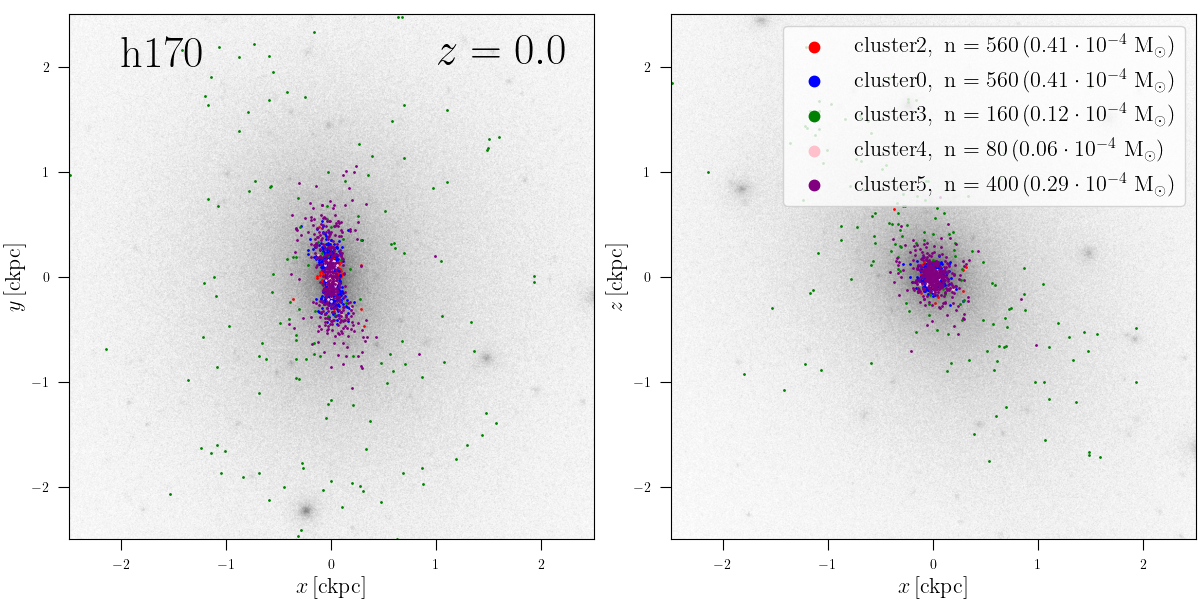}
  \includegraphics[width=0.39\textwidth]{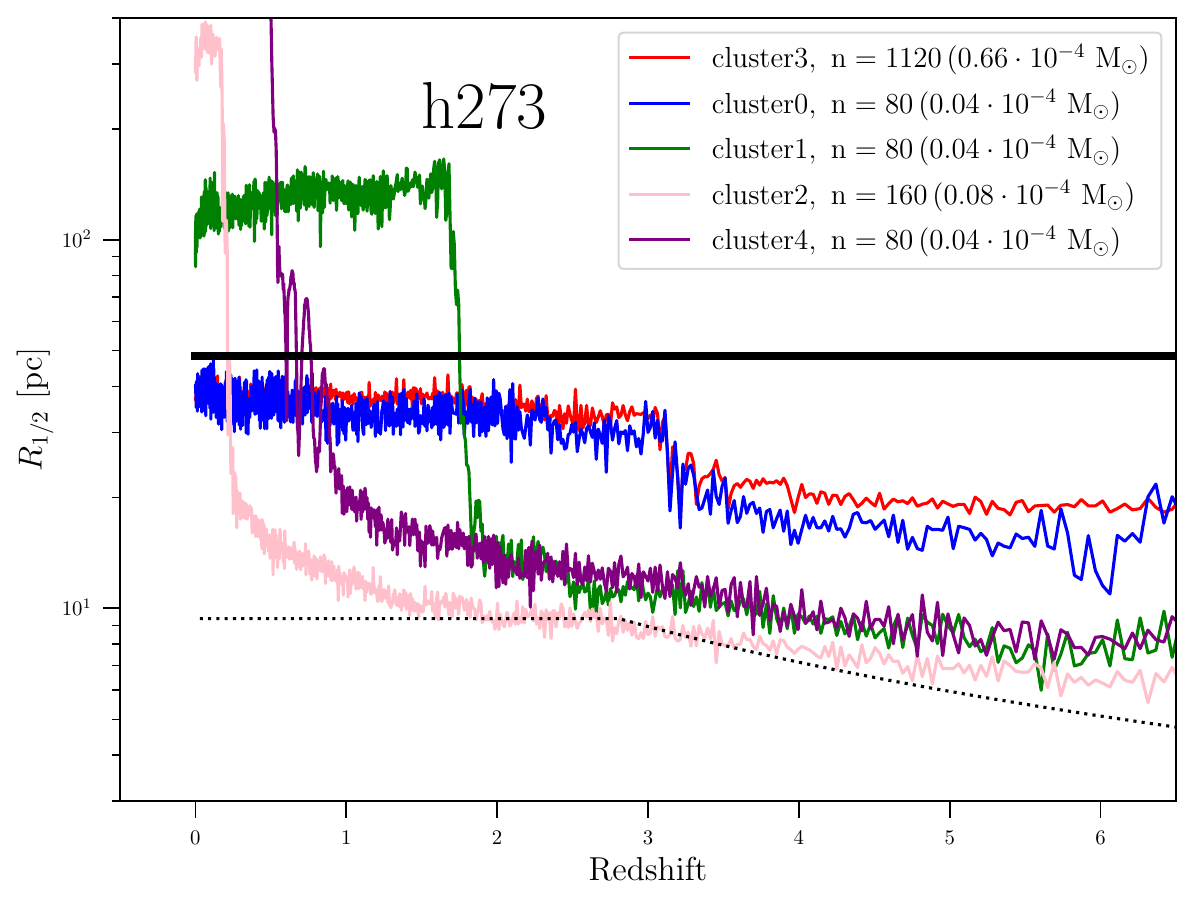}\includegraphics[width=0.60\textwidth]{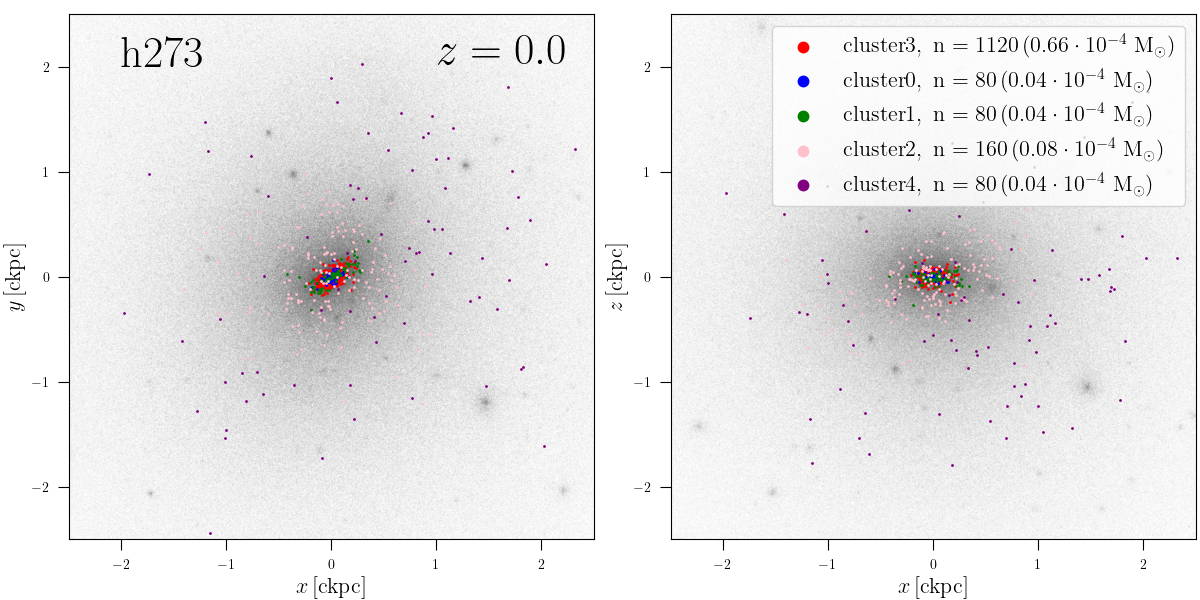}
  \includegraphics[width=0.39\textwidth]{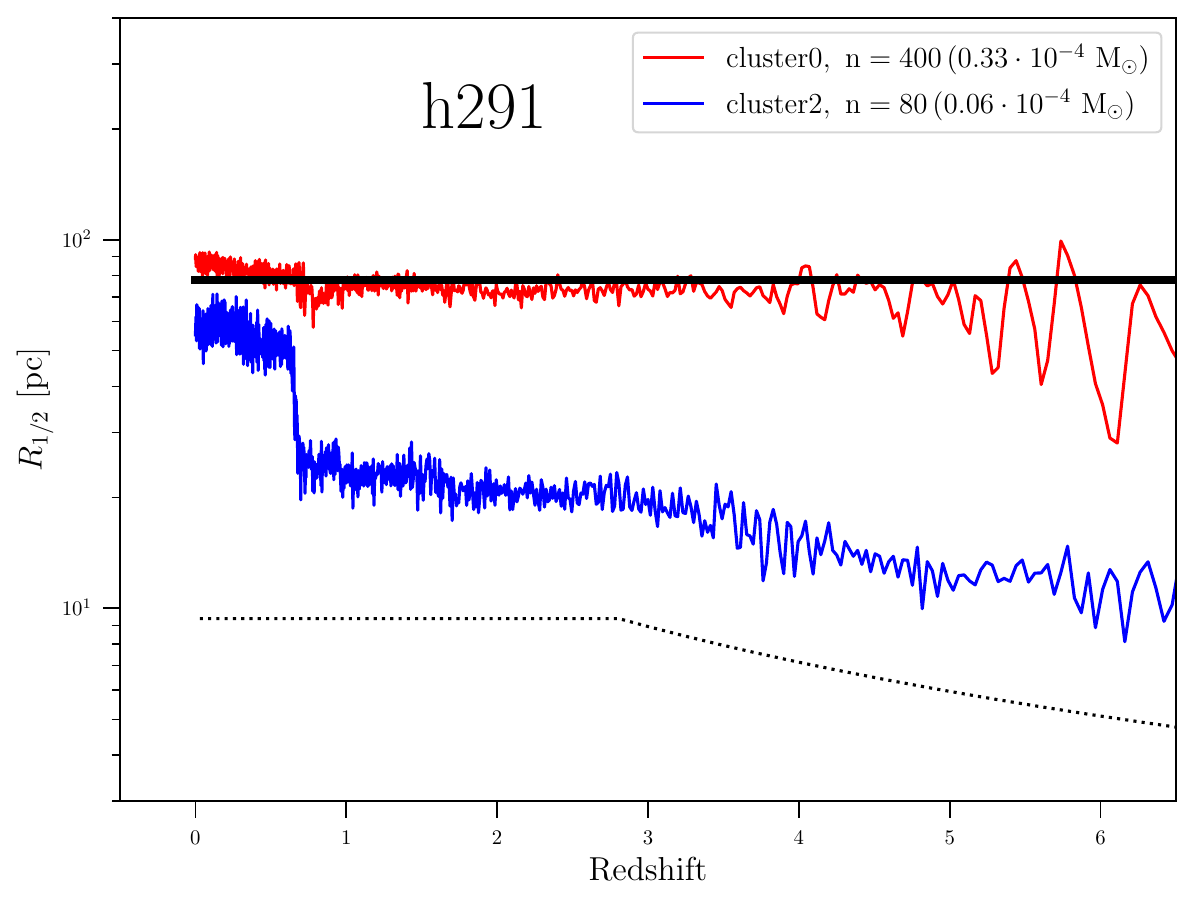}\includegraphics[width=0.60\textwidth]{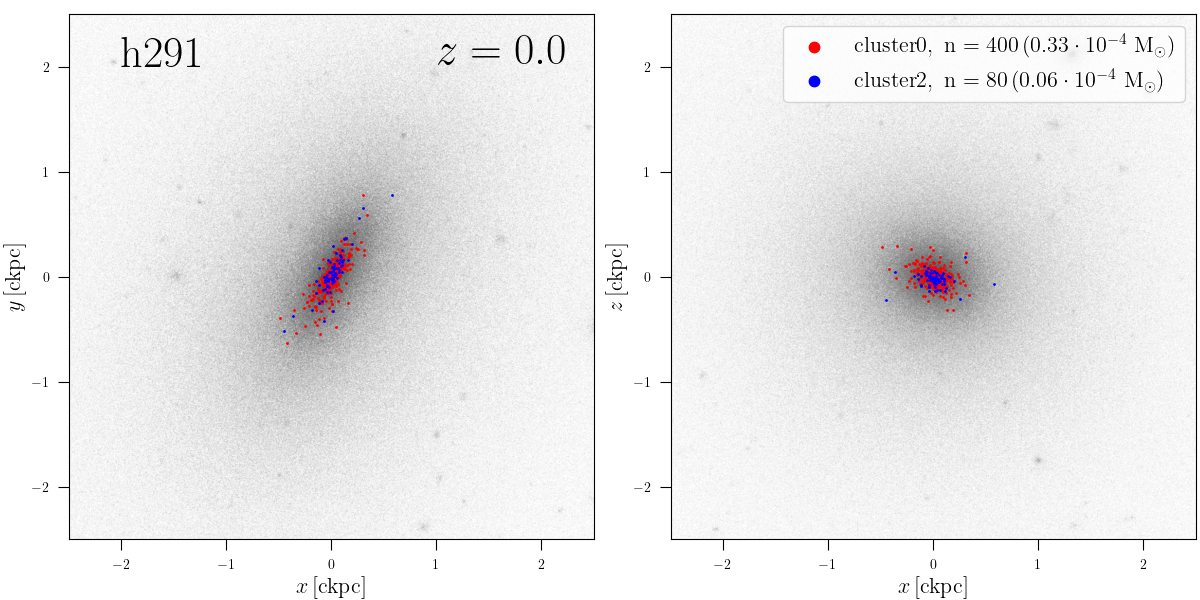}
  \end{subfigure}
  
  \caption{As in Fig.~\ref{fig:R12vsRedshift1} but for the three halos \texttt{h170},  \texttt{h273}, and \texttt{h291}.}
  
  \label{fig:R12vsRedshift2}

\end{figure*}

%%%%%%%%%%%%%%%%%%%%%%%%%%%%%%%%%%%%%%%%%%%%%%%%%%%%%%%%%%%%%%%%
\subsection{Properties at $z=0$}

The size and line-of-sight velocity dispersion of the stellar clusters at $z=0,$ 
and therefore those of the resulting UFDs, are given by the two last columns of Table~\ref{tab:ufds2}.
All halos but \texttt{h323} end up with a half mass of greater than $30\,\rm{pc}$, 
and none of them are able to reproduce the most compact UFD candidates 
reported in Fig.~\ref{fig:r12L}. 
The right panels of Figs.~\ref{fig:R12vsRedshift1} and \ref{fig:R12vsRedshift2} display 
two different projections of the clusters at the end of the simulation. 
These clusters are colour coded according to their colour attributed in the previous plots.
It is striking that, except for \texttt{h323}   again, where the dominant cluster does 
not merge with any of the others, all clusters exhibit complex morphological structures, 
which is the direct result of the assembly of those initially disconnected building blocks.
In \texttt{h277} and \texttt{h291,} the dominant clusters are strongly elongated.
On the contrary, in \texttt{h249} and \texttt{h170,} it is subdominant accreted 
clusters that form such an elongated structure. In  \texttt{h315} and \texttt{h273,} 
the destruction of accreted clusters leads to the formation of a stellar halo surrounding 
the main stellar system with an extension up to $2\,\rm{kpc}$, which is about 20 times the 
half-mass radius.

As mentioned in Sect. 1, UFDs are characterised by a mean ellipticity of $0.4$
\citep{martin2008} with some extreme cases showing values of greater than $0.6,$ 
as in Hercules and Bo\"otes I \citep{roderick2015,longeard2022}. 
While it is often claimed that such features are
evidence of gravitational tides generated by interaction with the Milky Way,
our simulations show that such morphological features could also result from the late
merger of clusters initially hosted in different subhalos: a natural prediction of a 
hierarchical cosmology.
Moreover, extended stellar structures 
have been observed in Tucana II \citep{chiti2021} with member stars found up to 9
half-light radii. Such an extended halo is reproduced here by the merger of clusters (
\texttt{h315} and \texttt{h273}), a scenario also supported by \citet{tarumi2021}.

We conclude that cluster mergers quickly lead to enlarged UFDs with extended and/or 
elongated stellar structures. Compact UFDs must result from a simple buildup history 
where the bulk of the stars form in one single mini-halo as observed for \texttt{h323}.

We finally compare the line-of-sight velocity dispersion of the stellar particles
in the final UFDs with those derived from observations.
The derived values range between $2.9$ and $4.4\,\rm{km/s}$,
in good agreement with those of UFDs \citep{Simon_2019}; except for
Hercules, for which a velocity dispersion as high as $8\,\rm{km/s}$ was recently derived \citep{longeard2023}.
We observe a clear correlation between $R_{1/2,\rm{fin}}$ and
$\sigma_{\rm{LOS,fin}}$, with the larger system being kinetically
warmer. Once more here, \texttt{h323} is an outlier as it is the only one with a
velocity dispersion below $3\,\rm{km/s}$, which is certainly the result of the absence of a merger.
%A similar correlation exists with the halo mass 

%dwarf dwarf merger \citep{deason2014}

%%%%%%%%%%%%%%%%%%%%%%%%%%%%%%%%%%%%%%%%%%%%%%%%%%%%%%%%%%%%%%%%
% hydro: l10 : gas=760, DM=4171 stars=430
% DMO  : l12 : DM=77 
\subsection{Influence of the number of particles per cluster}
\label{sec:fstar_influence}

Up to this point, the number of particles forming clusters is set using  
$f_{\star}=f_{\star,\rm{fid}}=32$, that is, we used 32 times more particles than the 
number of stars found per subhalo in the hydrodynamics run.
Following our particle selection method, that is, defining a cluster as a group formed by 
the most gravitationally bound particles, the number of particles considered for each 
subhalo at $z=6$ will determine the initial size of the cluster $R_{1/2,\rm{init}}$,
which could further impact the final size of the UFDs at $z=0$. 
Here, we explore the sensitivity of our previous results with respect to the value 
of $f_{\star}$.
More precisely, we use the ratio $f_{\star,\rm{fid}}/f_{\star}$, which we vary from 
1 to 16. Here, a value of 16 means that 16 times fewer particles are used  than in
our fiducial setup. Table~\ref{tab:content} provides a complete view of the size of all clusters 
at $z=6,$ as well as the final size of the UFDs at $z=0$ for different $f_{\star,
\rm{fid}}/f_{\star}$ ratios. We note that using a larger reduction factor is 
inadequate given our current resolution. The initial size of clusters would be
smaller than the gravitational softening length.

\begin{figure}
\includegraphics[width=0.49\textwidth]{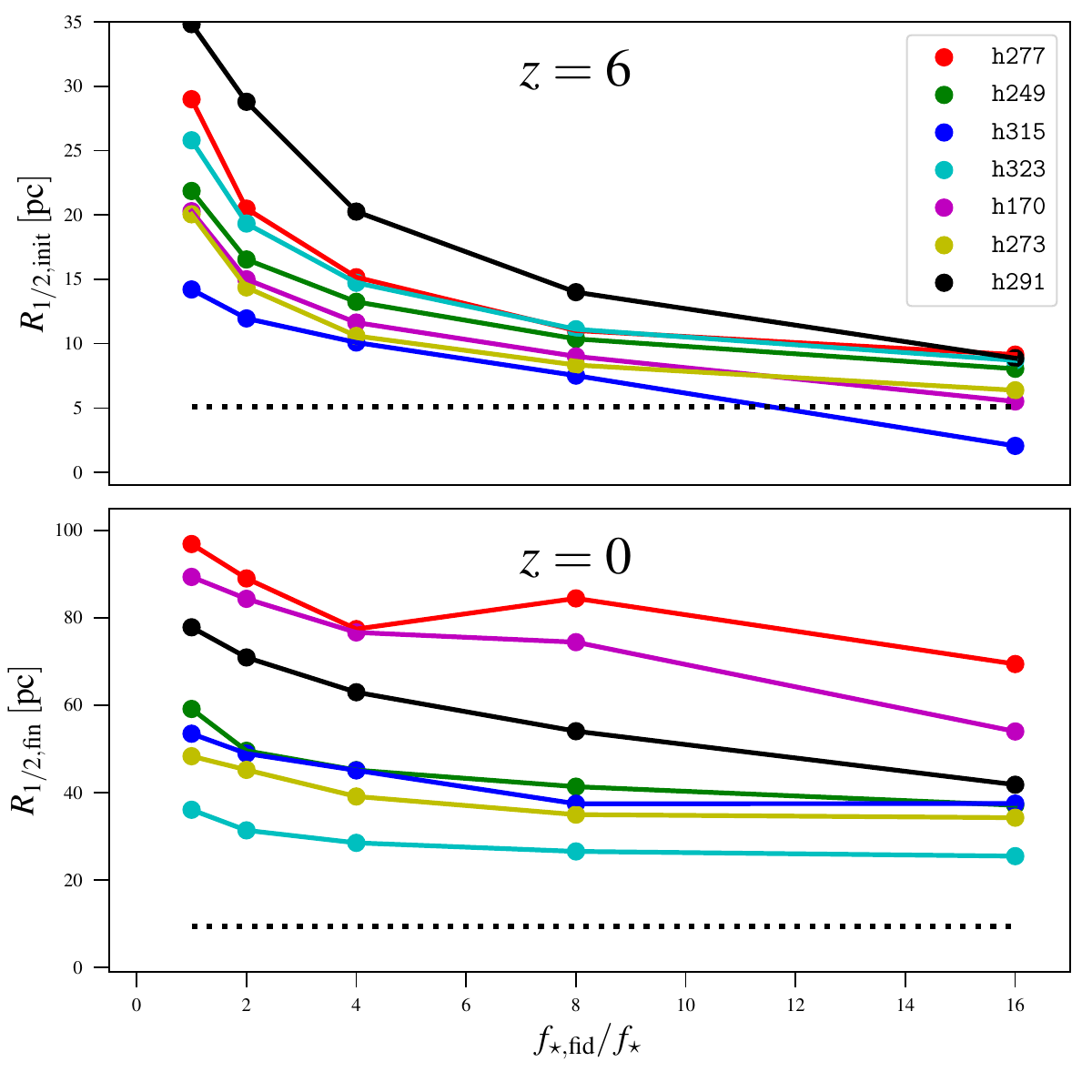}
\caption{Dependency of the initial size of the  main cluster $R_{1/2,\rm{init}}$ at $z=6$ (\emph{top panel}) and the final size of  UFDs  $R_{1/2}$ (\emph{bottom panel}) on the $f_\star$ parameter normalised here
by the fiducial value $f_{\star,\rm{fid}}$. Moving from 1 to 16, the number of particles in clusters is reduced
accordingly. In both panels, the dotted horizontal line indicates the simulation resolution, that is, 
the gravitational softening length.
}
\label{fig:fstar_convergence}
\end{figure}

For each of our seven halos,
Fig.~\ref{fig:fstar_convergence} shows both the initial size of the mass-dominant 
cluster ($R_{1/2,\rm{init}}$) and the size of the final UFD ($R_{1/2,\rm{fin}}$)  
with respect to $f_{\star,\rm{fid}}/f_{\star}$. 
As expected, reducing the number of particles in clusters leads to both decreasing
$R_{1/2,\rm{init}}$ and $R_{1/2,\rm{fin}}$.
If we ignore halo \texttt{h291} for a while, applying a reduction factor of 4
reduces the initial size of all main clusters down to about $10\,\rm{pc}$.
With this setting, we still guarantee that the main halos contain a 
substantial number of particles, that is, more than 100 (only \texttt{h315} is 
left with 80 particles in \texttt{cluster0}). 
Regarding the more compact systems (\texttt{h323}, \texttt{h273}, 
\texttt{h249}, \texttt{h315}), the curves corresponding to the size of the final UFDs 
flatten starting from 4 (bottom panel of Fig.~\ref{fig:fstar_convergence}), 
indicating a convergence. 
On average, we notice that decreasing the number of particles by a factor 16 leads
to a reduction of the initial size by a factor of $3.3$. However, this induces only a 
reduction of the final UFD size by  $1.3$ ($30\%$).
Importantly, at $z=0$, all clusters display 
a final size of greater than $20\,\rm{pc}$, independently  of the choice of $f_{\star}$.

%Reducing by a larger factor, 8 or 16, does not lead to a substantial reduction of 
%the cluster size. However the risk of a loss of accuracy owing to a low number of 
%particles increases.
%Not shown on Fig.~\ref{fig:fstar_convergence} is how realistic it is to consider
%clusters with 16 times less particles compared to the fiducial choice.

\begin{figure}
  \includegraphics[width=0.49\textwidth]{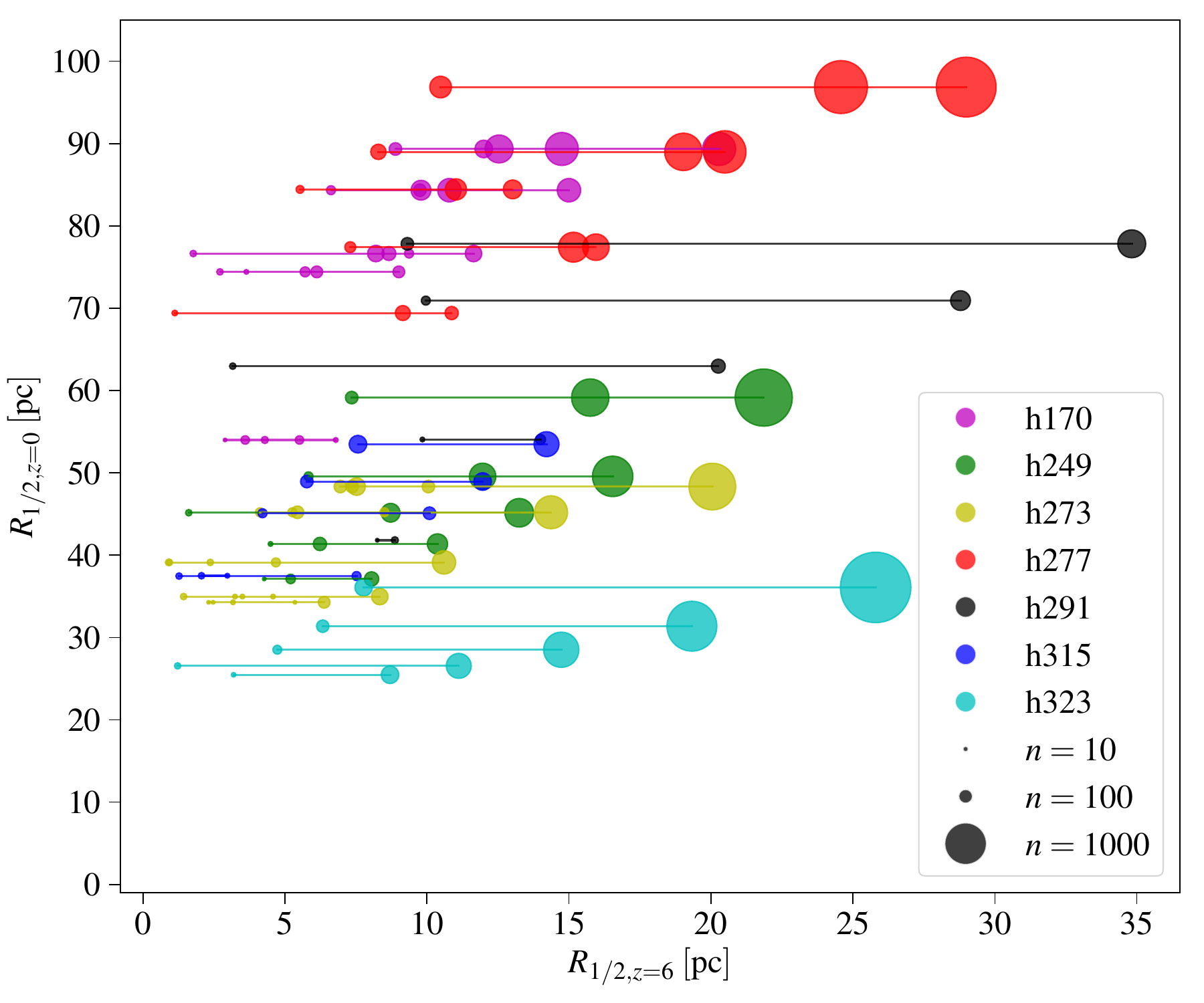}
\caption{Relation between the size of clusters at $z=6$ and the final size of the UFDs formed at $z=0$.
Symbol size scales with the number of particles defining the cluster.
Each UFD appears five times, for the five different $f_{\star,\rm{fid}}/f_{\star}$ ratios considered,
1, 2, 4, 8, or 16. Smaller UFDs correspond to larger $f_{\star,\rm{fid}}/f_{\star}$.
The detailed values for each cluster and UFD are provided in Table \ref{tab:content}.
}
\label{fig:R12finvsR12init}
\end{figure}

Figure~\ref{fig:R12finvsR12init} shows the relation between 
the size of each cluster at $z=6$ and the final size of the 
UFDs at $z=0$. Clusters forming the same UFD are connected by a thin line
and their size scales with the number of particles they contain.
Each UFD is represented by a different colour and for each of them, a different
$f_{\star}$ is shown through the $f_{\star,\rm{fid}}/f_{\star}$ ratio 1, 2, 4, 8, or 16.
Models with a specific ratio may be recognisable (counting from the top to the bottom), 
as a larger  $f_{\star,\rm{fid}}/f_{\star}$ leads to a smaller $R_{1/2}$.
% first
First and foremost,
Fig.~\ref{fig:R12finvsR12init} tells us that while starting with clusters
as compact as $10\,\rm{pc}$, the resulting UFD will have a half mass of greater than $30\,\rm{pc}$, and
up to $70\,\rm{pc}$ for model \texttt{h277}.
% and SO ?
UFDs that are built from more than one cluster, that is, all but 
\texttt{h323}, have a final $R_{1/2}$ that is more compact than $50\,\rm{pc}$ and require
the size of the initial cluster to be less than $20\,\rm{pc}$.
However, if UFDs are only formed by one single cluster (\texttt{h323}),
this constraint is slightly relaxed. An initial size of $20\,\rm{pc}$
is enough to get $R_{1/2}=32\,\rm{pc}$.
Only this model can be more compact than $30\,\rm{pc}$, 
requiring its dominant cluster to start with at most a size of $16\,\rm{pc}$. 
Finally, it is worth mentioning that any system that start larger will 
end up building a larger UFD.

%This is important for the conclusion when discussing 
%the impact of stellar physics.

%%
%\begin{figure}[h]
%  \centering
%  \includegraphics[width=0.48\textwidth]{graphs/UFDsR12vsLv/LR2.pdf}
%  \caption{\small Same figure than Fig.~\ref{fig:r12L} supplemented with the position of the current
%  model shown with the red stars.
%  %We extended our own work \citep{Revaz2018} with the results of unpublished galaxies in the UFDs regime.
%  \label{fig:r12L2}
%  }  
%\end{figure}
%%

%%%%%%%%%%%%%%%%%%%%%%%%%%%%%%%%%%%%%%%%%%%%%%%%%%%%%%%%%%%%%%%%
\subsection{Suppressing the formation of nearby clusters}

% impact of Ly-werner
As described in Sect.~\ref{sec:simulations}, our determination of the clusters
relies on the stellar content of dark matter halos predicted by \citet{sanati2023}. 
%As all hydro-dynamical cosmological simulations, those simulations face limitations either in term
%of resolution or implemented physics. 
One limitation to the physical model those simulations use is the lack of 
Lyman-Werner radiation from massive stars, which photo-dissociates the $\rm{H}_2$ molecule and strongly reduces the gas cooling. 
In a pristine interstellar medium, this process can suppress the formation of stars in the neighbourhood of 
the first formed building blocks \citep[e.g.][]{wise2009}.
While dedicated simulations with proper treatment of radiative transfer are needed
to  investigate whether or not this UV flux is strong enough to impact the formation site of the neighbouring 
physically decoupled clusters, our simulations can provide some estimations of the size of the resulting UFDs 
in an extreme case.
Assuming that all clusters surrounding the main cluster have failed to form,
the half-mass radius of a UFD will be that of its main cluster at $z=0$. 
Table~\ref{tab:lymanwerner} gives the final size  of each main cluster, together with an estimation
of its mass. To maximise the effect, we choose $f_{\star,\rm{fid}}/f_{\star}$ such that the clusters
have an initial size of about $10\,\rm{pc}$ (See~\ref{tab:content}).
This extreme setting helps to form more compat UFDs, and with the exception of model \texttt{h277}, which remains very extended, three UFDs with 
a half mass as low as $25\,\rm{pc}$ can be formed.

\begin{table}[h]
  \caption{\small Final size of the main clusters excluding the contribution of all other clusters.}
    \label{tab:lymanwerner}
  \centering
  %\tiny
  \begin{tabular}{l c c c c c c }
  \hline
  \hline
    Halo  & Cluster  & $f_{\star,\rm{fid}}/f_{\star}$ & \# & $M\star$ & $R_{1/2,\rm{ini}}$ & $R_{1/2,\rm{fin}}$ \\
    ID        &     ID       &      & &         $[10^4\,\rm{M_\odot}]$                   & $[\rm{pc}]$   & $[\rm{pc}]$      \\
  \hline
  \hline
  \texttt{h277} & 1 & 8 & 230 & 0.170&11.02  & 185.14\\
  \texttt{h249} & 2 & 8 & 201 & 0.148&10.37  & 24.97\\
  \texttt{h315} & 0 & 4 &  80 & 0.059&10.09  & 29.15\\
  \texttt{h323} & 0 & 8 & 319 & 0.235&11.12  & 26.26\\
  \texttt{h170} & 2 & 4 & 140 & 0.103&11.64  & 25.72\\
  \texttt{h273} & 3 & 4 & 280 & 0.207&10.60  & 32.09\\
  \texttt{h291} & 0 & 8 &  50 & 0.037&14.00  & 43.99\\
  \hline
  \hline
  \end{tabular}%

  \tablefoot{
  $f_{\star,\rm{fid}}/f_{\star}$ is chosen such that the initial
  size of the main cluster  is about $10\,\rm{pc}$.
  }

\end{table}

%%%%%%%%%%%%%%%%%%%%%%%%%%%%%%%%%%%%%%%%%%%%%%%%%%%%%%%%%%%%%%%%

\section{Discussion and Conclusions}\label{sec:conclusions}

%%%%%%%%%%%%%%%%%%%%%%%%%%%%%%%%%%%%%%%%%%%%%%%%%%%%%%%%%%%%%%%%
%

Motivated to understand the difficulty cosmological hydrodynamics simulations have in
reproducing the compactness of observed UFDs, we performed a set of seven very high-resolution DMO simulations of UFDs. Using a simple but powerful and numerically clean 
method that combines predictions of the stellar content of UFDs from hydrodynamics simulations 
with those new DMO runs, we followed the buildup history of UFDs; that is, how 
stellar building blocks (referred to here as clusters) formed before the end of the EoR assemble to form a UFD at $z=0$. We focused in particular on the evolution of the size
of those clusters as well as on the size of the resulting UFDs.

%We emphasize that in our approach, all particles share the same mass. We thus avoid numerical problems that can lead to a spurious dynamical heating and and over-estimation of the size of the systems \citep{Revaz2018,ludlow2019,ludlow2021}.

Our main findings are summarised below:

\begin{itemize}

\item At $z=6$ for all models, the stellar building blocks that 
assemble into a UFD at $z=0$ are spatially disconnected; that is, they are hosted in 
separate mini-halos with relative distances that can extend to $130\,\rm{ckpc}$ 
(e.g. \texttt{h291}).

\item The fusion of these disconnected mini-halos has a strong impact on the
final size of the resulting UFDs. From redshift 6 down to 0, we observe a general
increase in the size of the stellar systems. This increase is the 
direct impact of cluster--cluster mergers including their respective host halo,
or mergers between clusters and star-free halos.
%merger, where the halo can be star-free.

\item As a consequence of these mergers, none of our experiments
lead to the formation of a UFD with a half mass (half light) 
smaller than $20\,\rm{pc}$, that is, regardless of $f_{\star}$, the parameter that 
sets the number of stars per mini-halo.

\item Ultra-faint dwarf galaxies, with a size smaller than $30\,\rm{pc}$ are obtained only 
if all stars are formed in one single mini-halo at $z=6$,
as in \texttt{h323}. The associated halo mass must be smaller than $4\cdot 10^8\,\rm{M_\odot}$
and the size of the initial cluster smaller than $15\,\rm{pc}$.
More massive halos can form similarly compact UFDs, but only if the formation of neighbouring 
halos is prevented by, for example, Lyman-Werner radiation.
In both cases, at $z=6$, the stellar content must be strongly gravitationally bound.

\item Forming  UFDs with a size of between $30$ and $50\,\rm{pc}$ is possible 
despite more complex buildup histories, that is if the initial cluster size is smaller 
than $20\,\rm{pc}$ (\texttt{h249}, \texttt{h273}, \texttt{h315}) and
all stellar particles are the most gravitationally bound ones.

\item If a larger cluster forms, for example with a size of $35\,\rm{pc}$,
the size of the UFD ends up relatively large ($\cong 80\,\rm{pc}$), even in the case of a simple buildup involving two clusters (\texttt{h291}).

\item A UFD that forms via a complex buildup involving several initially distinct 
clusters will be naturally extended, with $r_{1/2}\gtrapprox 100\,\rm{pc}$ 
(\texttt{h170}, \texttt{h277}), despite the stellar building blocks being initially 
very compact and  strongly gravitationally bound.

\item The fusion of initially spatially dissociated clusters 
systematically leads to peculiar signatures imprinted in the UFD stellar 
component, such as a pronounced ellipticity.
Those structures could easily be confused with tidal tails 
\citep{martin2008,aden2009,roderick2015,longeard2022}.
As in \citet{tarumi2021}, we also observed the formation of extended stellar halos
in three of our models (\texttt{h315}, \texttt{h170}, \texttt{h273}).
Such a feature is in line with the extended stellar population found around 
Tucana II \citep{chiti2021}. It also supports the finding  by  \citet{frebel2016} that multiple stellar populations
observed in Bo\"otes can originate from different 
stellar building blocks.

%\item Even if resulting from the fusion of initially very compact clusters 
%($R_{1/2}<7\,\rm{pc}$), \texttt{h273} can never be more compact than 
%$70\,\rm{pc}$ owing to it slightly larger virial mass $6\cdot 10^8\,\rm{M_\odot}$
%that leads to a complex buildup history that involve five clusters.

\end{itemize}

%Cetus2        0.014 no ufd
%Draco2        0.019  #Fu2023
%Eridanus3     0.007  #Fu2023
%VirgoII       0.016  ?
%DESJ0225+0304 0.018  ?

%Indus1        0.025  ?
%Pictor1       0.022  ?
%Segue(1)      0.026
%Willman1      0.028
%BoootesV      0.020
%LeoMinorI     0.026

The above conclusions lead us to the following more general conclusions and comments.
The difficulties in reproducing the most compact UFDs 
$(R_{1/2}<20\,\rm{pc})$ must be taken with care.
So far, no velocity dispersion 
measurements have been obtained to confirm the true nature of any of the four UFD candidates observed
which such a small size.
However, for two of them, Eridanus III and Draco II, determination of their metallicity distribution function was recently obtained with the HST CaHK narrow band \citep{fu2023}.
Both show a wide range of metallicities, supporting their UFD nature.

% R12 < 30 -> contraints on the feedback
Forming very compact UFDs ($R_{1/2}<30\,\rm{pc}$) requires that all member stars 
originate from the same mini-halo prior to the end of the EoR. 
In the absence of Lyman-Werner radiation,
such a requirement imposes that UFDs form in relatively light
dark halos with a virial mass of less than ($3\cdot 10^8\,\rm{M_\odot}$).
This seems to be in agreement with the results of \citet{jeon2021a}, where
the most compact UFDs were found ---in hydrodynamical simulations 
\footnote{However, see our remark in Sect.~\ref{sec:challenge} regarding the resolution 
of those simulations.}---  to be reproduced within halos lighter than 
$1.5\cdot 10^8\,\rm{M_\odot}$.
About $50$ to $100$ such halos are expected within 
$300\,\rm{kpc}$ around the Milky Way \citep{sawala2017},
in agreement with predictions of the observed luminosity 
function of Milky Way satellites \citep{tollerud2008,newton2018,nadler2019}.

% R12 < 50 -> contraints on the feedback
Compact UFDs with $R_{1/2}<50\,\rm{pc}$ can be formed in a hierarchical 
cosmology but remain challenging as they require the size of the initial 
stellar building blocks to be more compact than $20\,\rm{pc}$ at $z=6$.
While this is achieved in the present DMO simulations, we recall that
we define clusters as the most gravitationally bound particles in our systems,
and therefore, by construction, neglected any physical process that could lead to their 
pre-heating and thus increase their size
at the end of the EoR. In this sense, our method is extremely conservative.
Any additional baryonic effect, such as stellar feedback, should undoubtedly lead to a more extended 
stellar system  than the ones obtained here, possibly maintaining 
some tension with the data.

% lack of interaction
All simulations presented here were performed in a cosmological 
context, although without including tidal effects induced by the presence of 
a gravitationally dominating Milky Way. How would these effects impact the 
results ? The smallest halos could be expected to be 
disrupted, thus reducing cluster--cluster mergers and preventing the size increase of 
the survivors.
However, it is important to underline that the clusters followed in this 
work are deeply embedded in massive dark halos extended up to $20\,\rm{kpc}$, which is 
roughly three orders of magnitude larger than the size of the stellar component. 
Those dark halos can act as a gravitational shield \citep{penarrubia2008}, 
preventing the disruption of their stellar counterpart. 
Dedicated simulations will be necessary to answer this point.

We also emphasise that the difficulty in reproducing the compactness of UFDs originates
from the hierarchical formation of structures at the smallest scales, which is imposed by 
CDM. By suppressing the smallest dark halos, that is, not only their stellar content 
(potentially through $H_2$ dissociation) but also the halos themselves, 
one can potentially reduce this tension.
Such suppression is a direct expectation from warm dark matter (WDM) models, which are worthy of further study.

% importance for chemical evolution
Finally, from a chemical evolution point of view, we underline the importance of the 
complex buildup of the more massive UFDs assembled from several initially 
physically decoupled stellar clusters. 
Putting together stellar populations that evolved separately for a certain amount of time can 
lead to the formation of a stellar system with inhomogeneous abundances; 
such inhomogeneities could therefore be used as a tracer of hierarchical assembly at the smallest scales.
%an indicator of a hierarchical assembly
%as expected from the $\Lambda$CDM paradigm.

%extremely diffuse UFDs \citep{rey2019}

%
%%%%%%%%%%%%%%%%%%%%%%%%%%%%%%%%%%%%%%%%%%%%%%%%%%%%%%%%%%%%%%%%

\begin{acknowledgements}

%%%%%%%%%%%%%%%%%%%%%%%%%%%%%%%%%%%%%%%%%%%%%%%%%%%%%%%%%%%%%%%%
%

We thank the referee for the very positive and constructive report, which
helped to clarify and improve the manuscript. 
We thanks Nathan Monnet for helping in the identification of halo pairs.
We are grateful to Myoungwon Jeon, Nicolas Longeard, Pascale Jablonka, Mahsa Sanati,
Mladen Ivkovic for stimulating discussions.
The data reduction and surface density maps have been performed using the parallelised Python
\textsc{pNbody} package \citep{revaz2013}\footnote{\url{https://lastro.epfl.ch/projects/pNbody/}}.
\end{acknowledgements}

%
%%%%%%%%%%%%%%%%%%%%%%%%%%%%%%%%%%%%%%%%%%%%%%%%%%%%%%%%%%%%%%%%
\bibliographystyle{aa}
\bibliography{bibliography}
%%%%%%%%%%%%%%%%%%%%%%%%%%%%%%%%%%%%%%%%%%%%%%%%%%%%%%%%%%%%%%%%
%
\clearpage
\onecolumn

%
%%%%%%%%%%%%%%%%%%%%%%%%%%%%%%%%%%%%%%%%%%%%%%%%%%%%%%%%%%%%%%%%

\begin{appendix}

%%%%%%%%%%%%%%%%%%%%%%%%%%%%%%%%%%%%%%%%%%%%%%%%%%%%%%%%%%%%%%%%
%

\section{Detailed properties of extracted clusters and UFDs.}

\begin{table*}[h]
  \caption{\small Details of the number of particles ($N$) and size ($R_{1/2}$) of all clusters at $z=6$ 
  and the resulting UFDs at $z=0$.}
    \label{tab:content}
  \centering
  %\tiny

  \begin{tabular}{l l r r r r r r r r r r }
  \hline
  \hline
  \\
    & &
    \multicolumn{2}{c}{$f_{\star,\rm{fid}}/f_{\star}=1$} &
    \multicolumn{2}{c}{$f_{\star,\rm{fid}}/f_{\star}=2$} &
    \multicolumn{2}{c}{$f_{\star,\rm{fid}}/f_{\star}=4$} &
    \multicolumn{2}{c}{$f_{\star,\rm{fid}}/f_{\star}=8$} &
    \multicolumn{2}{c}{$f_{\star,\rm{fid}}/f_{\star}=16$} \\\\

    Halo ID & 
    Cluster ID &
    $R_{1/2}/\rm{[pc]}$ &
    N &
    $R_{1/2}/\rm{[pc]}$ &
    N &
    $R_{1/2}/\rm{[pc]}$ &
    N &    
    $R_{1/2}/\rm{[pc]}$ &
    N &
    $R_{1/2}/\rm{[pc]}$ &
    N \\\\    
    
  \hline
  \\
\texttt{h277}  &  \rm{cluster1}  &  $28.99$  &  $1840$  &  $20.49$  &  $ 920$  &  $15.16$  &  $460$  &  $11.02$  &  $230$  &  $ 9.15$  &  $115$ \\
               &  \rm{cluster0}  &  $24.58$  &  $1440$  &  $19.03$  &  $ 720$  &  $15.95$  &  $360$  &  $13.02$  &  $180$  &  $10.87$  &  $ 90$ \\
               &  \rm{cluster3}  &  $10.48$  &  $ 240$  &  $ 8.29$  &  $ 120$  &  $ 7.30$  &  $ 60$  &  $ 5.53$  &  $ 30$  &  $ 1.12$  &  $ 15$ \\\\
               &  \rm{UFD}       &  $96.86$  &  $3520$  &  $88.99$  &  $1760$  &  $77.41$  &  $880$  &  $84.43$  &  $440$  &  $69.41$  &  $220$ \\\\
\texttt{h249}  &  \rm{cluster2}  &  $21.86$  &  $1680$  &  $16.54$  &  $ 840$  &  $13.25$  &  $420$  &  $10.37$  &  $210$  &  $ 8.05$  &  $105$ \\
               &  \rm{cluster0}  &  $15.75$  &  $ 720$  &  $11.96$  &  $ 360$  &  $ 8.72$  &  $180$  &  $ 6.23$  &  $ 90$  &  $ 5.20$  &  $ 45$ \\
               &  \rm{cluster1}  &  $ 7.35$  &  $  80$  &  $ 5.83$  &  $  40$  &  $ 1.61$  &  $ 20$  &  $ 4.49$  &  $ 10$  &  $ 4.27$  &  $  5$ \\\\
               &  \rm{UFD}       &  $59.14$  &  $2480$  &  $49.57$  &  $1240$  &  $45.15$  &  $620$  &  $41.36$  &  $310$  &  $37.11$  &  $155$ \\\\
\texttt{h315}  &  \rm{cluster0}  &  $14.21$  &  $ 320$  &  $11.96$  &  $ 160$  &  $10.09$  &  $ 80$  &  $ 7.52$  &  $ 40$  &  $ 2.06$  &  $ 20$ \\
               &  \rm{cluster1}  &  $ 7.57$  &  $ 160$  &  $ 5.77$  &  $  80$  &  $ 4.21$  &  $ 40$  &  $ 1.27$  &  $ 20$  &  $ 2.97$  &  $ 10$ \\\\
               &  \rm{UFD}       &  $53.48$  &  $ 480$  &  $48.94$  &  $ 240$  &  $45.09$  &  $120$  &  $37.46$  &  $ 60$  &  $37.50$  &  $ 30$ \\\\
\texttt{h323}  &  \rm{cluster0}  &  $25.80$  &  $2556$  &  $19.33$  &  $1278$  &  $14.73$  &  $639$  &  $11.12$  &  $319$  &  $ 8.70$  &  $159$ \\
               &  \rm{cluster1}  &  $ 7.77$  &  $ 156$  &  $ 6.33$  &  $  78$  &  $ 4.73$  &  $ 39$  &  $ 1.22$  &  $ 19$  &  $ 3.19$  &  $  9$ \\\\
               &  \rm{UFD}       &  $36.08$  &  $2712$  &  $31.37$  &  $1356$  &  $28.51$  &  $678$  &  $26.55$  &  $338$  &  $25.47$  &  $168$\\\\
\texttt{h170}  &  \rm{cluster2}  &  $20.29$  &  $ 560$  &  $15.00$  &  $ 280$  &  $11.64$  &  $140$  &  $ 9.01$  &  $ 70$  &  $ 5.51$  &  $ 35$ \\
               &  \rm{cluster0}  &  $14.75$  &  $ 560$  &  $10.79$  &  $ 280$  &  $ 8.20$  &  $140$  &  $ 6.12$  &  $ 70$  &  $ 3.60$  &  $ 35$ \\
               &  \rm{cluster3}  &  $12.00$  &  $ 160$  &  $ 9.75$  &  $  80$  &  $ 9.37$  &  $ 40$  &  $ 2.71$  &  $ 20$  &  $ 6.79$  &  $ 10$ \\
               &  \rm{cluster4}  &  $ 8.89$  &  $  80$  &  $ 6.62$  &  $  40$  &  $ 1.77$  &  $ 20$  &  $ 3.64$  &  $ 10$  &  $ 2.89$  &  $  5$ \\
               &  \rm{cluster5}  &  $12.54$  &  $ 400$  &  $ 9.79$  &  $ 200$  &  $ 8.66$  &  $100$  &  $ 5.71$  &  $ 50$  &  $ 4.29$  &  $ 25$ \\\\
               &  \rm{UFD}       &  $89.34$  &  $1760$  &  $84.33$  &  $ 880$  &  $76.64$  &  $440$  &  $74.41$  &  $220$  &  $53.99$  &  $110$ \\\\
\texttt{h273}  &  \rm{cluster3}  &  $20.05$  &  $1120$  &  $14.37$  &  $ 560$  &  $10.60$  &  $280$  &  $ 8.34$  &  $140$  &  $ 6.38$  &  $ 70$ \\
               &  \rm{cluster0}  &  $10.05$  &  $  80$  &  $ 8.50$  &  $  40$  &  $ 2.37$  &  $ 20$  &  $ 4.58$  &  $ 10$  &  $ 5.35$  &  $  5$ \\
               &  \rm{cluster1}  &  $ 7.36$  &  $  80$  &  $ 5.26$  &  $  40$  &  $ 0.90$  &  $ 20$  &  $ 3.24$  &  $ 10$  &  $ 2.47$  &  $  5$ \\
               &  \rm{cluster2}  &  $ 7.52$  &  $ 160$  &  $ 5.44$  &  $  80$  &  $ 4.68$  &  $ 40$  &  $ 1.43$  &  $ 20$  &  $ 3.17$  &  $ 10$ \\
               &  \rm{cluster4}  &  $ 6.95$  &  $  80$  &  $ 4.12$  &  $  40$  &  $ 0.93$  &  $ 20$  &  $ 3.50$  &  $ 10$  &  $ 2.31$  &  $  5$ \\\\
               &  \rm{UFD}       &  $48.33$  &  $1520$  &  $45.21$  &  $ 760$  &  $39.12$  &  $380$  &  $34.97$  &  $190$  &  $34.27$  &  $ 95$ \\\\
\texttt{h291}  &  \rm{cluster0}  &  $34.82$  &  $ 400$  &  $28.79$  &  $ 200$  &  $20.26$  &  $100$  &  $14.00$  &  $ 50$  &  $ 8.87$  &  $ 25$ \\
               &  \rm{cluster2}  &  $ 9.31$  &  $  80$  &  $ 9.96$  &  $  40$  &  $ 3.16$  &  $ 20$  &  $ 9.84$  &  $ 10$  &  $ 8.25$  &  $  5$ \\\\
               &  \rm{UFD}       &  $77.82$  &  $ 480$  &  $70.92$  &  $ 240$  &  $62.96$  &  $120$  &  $54.05$  &  $ 60$  &  $41.81$  &  $ 30$ \\\\
  \hline
  \hline
  \end{tabular}%

  \tablefoot{
  Values are given for a specific $f_{\star,\rm{fid}}/f_{\star}$ ratio.
  }

\end{table*}

\end{appendix}

%%%%%%%%%%%%%%%%%%%%%%%%%%%%%%%%%%%%%%%%%%%%%%%%%%%%%%%%%%%%%%%%
\end{document}